\documentclass[lettersize,journal]{IEEEtran}
\usepackage{cite}
\usepackage{graphicx}
\usepackage{epstopdf}
\usepackage{url}
\usepackage[cmex10]{amsmath}
\usepackage{amsmath,amsthm,amsfonts,amssymb,latexsym,extarrows,booktabs,array}
\usepackage{subfigure}
\usepackage{threeparttable}
\usepackage{mathrsfs}
\usepackage[mathscr]{euscript}
\usepackage{multicol}
\usepackage{stfloats}
\usepackage{algorithm}
\usepackage{diagbox}
\usepackage{algorithmic}
\usepackage{accents}
\usepackage{xcolor}
\usepackage{lipsum,amsmath}
\usepackage{cuted}
\usepackage{bm}
\usepackage{times}
\usepackage[algo2e]{algorithm2e}
\usepackage{multirow}
\usepackage{makecell}
\usepackage{color}
\theoremstyle{definition}

\theoremstyle{remark}\newtheorem{Remark}{Remark}

\begin{document}

	\title{ {Stacked Intelligent Metasurface Assisted Multiuser Communications: From a Rate Fairness Perspective}}
	\bstctlcite{BSTcontrol}
	\author{
		Junjie~Fang,~Chao~Zhang,~Jiancheng~An,~Hongwen~Yu, and~Qingqing~Wu,~\IEEEmembership{Senior Member,~IEEE},
		\\~Mérouane~Debbah,~\IEEEmembership{Fellow,~IEEE}, and~Chau~Yuen,~\IEEEmembership{Fellow,~IEEE}
		\thanks{Junjie Fang and Chao Zhang are with School of Information and Communications Engineering, Xi'an Jiaotong University, Xi'an 710049, China (e-mail: fjjzcy@stu.xjtu.edu.cn, chaozhang@xjtu.edu.cn) (\emph{Corresponding author: Chao Zhang})}
		\thanks{Jiancheng An and Chau Yuen are with the School of Electrical and Electronics Engineering, Nanyang Technological University, Singapore 639798 (e-mail: jiancheng$\_$an@163.com, chau.yuen@ntu.edu.sg)}
		\thanks{Hongwen Yu is with the School of Communication and Information Engineering, Shanghai University, Shanghai 200444, China (e-mail: hw$\_$yu@shu.edu.cn)}
		\thanks{Qingqing Wu is with Department of Electronic Engineering, Shanghai Jiao tong University, Shanghai 200240, China (e-mail:  qingqingwu@sjtu.edu.cn).}
		\thanks{Mérouane Debbah is with Khalifa University of Science and Technology, P O Box 127788, Abu Dhabi, UAE (e-mail: merouane.debbah@ku.ac.ae)}
	}
	
	\markboth{}%
	{}

	\maketitle
	
\begin{abstract}
Stacked intelligent metasurface (SIM) extends the concept of single-layer reconfigurable holographic surfaces (RHS) by incorporating a multi-layered structure, thereby providing enhanced control over electromagnetic wave propagation and improved signal processing capabilities.  This study investigates the potential of SIM in enhancing the rate fairness in multiuser downlink systems by addressing two key optimization problems:  maximizing the minimum rate (MR) and maximizing the geometric mean of rates (GMR).  {The former strives to enhance the minimum user rate, thereby ensuring fairness among users, while the latter relaxes fairness requirements to strike a better trade-off between user fairness and system sum-rate (SR).} For the MR maximization, we adopt a consensus alternating direction method of multipliers (ADMM)-based approach, which decomposes the approximated problem into sub-problems with closed-form solutions.   {For GMR maximization, we develop an alternating optimization (AO)-based algorithm that also yields closed-form solutions and can be seamlessly adapted for SR maximization. Numerical results validate the effectiveness and convergence of the proposed algorithms.} Comparative evaluations show that MR maximization ensures near-perfect fairness, while GMR maximization balances fairness and system SR.    Furthermore,  the two proposed algorithms respectively outperform existing related works in terms of MR and SR performance. Lastly, SIM with lower power consumption achieves performance comparable to that of multi-antenna digital beamforming.
\end{abstract}
	
	\section{Introduction}

The sixth generation (6G) mobile communication network is envisioned to support peak data rates of up to 1 terabit per second and ultra-low latency of around 1 millisecond, driven by the demands of applications such as ultra-high-definition streaming, augmented and virtual reality (AR/VR), and the expanding Internet of Things (IoT)~\cite{Wang2023}.  As massive multiple-input multiple-output (MIMO) technology can provide significant spatial diversity and enhance network capacity greatly, it has been regarded as a promising technology to address these demands~\cite{Wang2023}. Unfortunately, it also presents challenges, including high implementation costs, substantial power consumption, and the complexity of real-time digital signal processing \cite{Wu2024}. Hence, as data demands continue to rise, there is an urgent need for more cost-effective and energy-efficient solutions  \cite{Wu2021}. Inspired by the robust spatial multiplexing and beamforming capability brought by massive MIMO, researchers are now exploring extremely dense (and even theoretically infinite) low cost arrays to achieve even greater spatial resolution and more precise beam control \cite{Huang2020}.

Emerging from breakthroughs in metasurfaces and intelligent material design, reconfigurable intelligent surface (RIS) is  perceived as a transformative technology in wireless communication systems due to a variety of fascinating features, such as low cost, low energy-consumption, low latency, high energy efficiency, and dynamic programmability\cite{Wu2024}. Specifically, a RIS typically consists of dense arrays of sub-wavelength passive/active elements that can dynamically adjust the phases and/or amplitudes of incident electromagnetic (EM) waves, thereby achieving remarkable fine-grained beamforming gains\cite{Wu2023a}. Currently, RIS can be roughly categorized into two groups based on their function in wireless communication systems \cite{Wei2022}. The first category, commonly referred to as intelligent reflecting surfaces (IRS), provides additional reflective links for transceivers to enhance link quality, extend coverage, and improve network capacity\cite{Wu2019}. The second category, commonly known as reconfigurable holographic surface (RHS), is typically integrated with radio frequency (RF) circuits and serves as  transceiver antenna\cite{Wei2022}. However, recent studies have shown that existing RHS is constrained by its single-layer  structure, which may limit the design degrees-of-freedom (DoF) for holographic beamforming\cite{Liu2022,An2023,An2023a,An2024,An2024a}.

A novel structure known as stacked intelligent metasurface (SIM) has recently gained increasing attention\cite{An2023a}. The pioneer work of SIM in \cite{Liu2022} demonstrated a programmable multi-layer digital-coding metasurface array, which is proved to be  suitable for signal processing in wireless systems. Further, in  \cite{An2023a}, An et al. proposed an SIM-assisted holographic MIMO framework, where an SIM consists of multiple layers of programmable metasurfaces housed in a hermetically sealed container, with sub-wavelength passive elements in each layer.  
The authors also affirmed that almost perfect parallel channels can be constructed between transmitter and receiver as the number of SIM layers increases. More specifically speaking,  the multi-layer structure allows for finer granularity in phase and amplitude modulation and provides increased flexibility in beam pattern adjustment.  To minimize processing delays and energy consumption associated with digital beamforming, as well as to mitigate the effects of two-hop path loss, SIM is typically integrated with transmit or/and receive antennas, which enables it to perform holographic precoding or/and combining directly in the EM wave domain \cite{An2025}. 

\begin{table*}[!t]
	\setlength\tabcolsep{5pt}
	\caption{Comparison of our work with existing SIM-related efforts}\label{tab:LiteratureComparison}
	\centering 
	{
		\begin{tabular}{|c|c|c|c|c|c|c|c|c| }
			\hline
			\multirow{2}{*}{Literature} & \multirow{2}{*}{MIMO Setup}  & \multicolumn{4}{c|}{Objective}  & \multirow{2}{*}{Method} & \multirow{2}{*}{Closed-form } & \multirow{2}{*}{Rate fairness}
			\\
			\cline{3-6}
			& & MR  & GMR  & Sum/Achievable rate  & MSINR  & &  &
			\\
			\hline
			\cite{An2023,An2025,Lin2024,Papazafeiropoulos2024a} & MU-MISO  &   &   & \checkmark  &    &  AO+PGA &  &
			\\
			\hline
			\cite{Liu2025} & MU-MISO  &   &   & \checkmark  &    & DRL & &
			\\
			\hline
			\cite{Kavianinia2025} & Secure MU-MISO  &   &   & \checkmark  & & AO+PGA  &  &
			\\
			\hline
			\cite{Li2024a} & ISAC  &   &   & \checkmark  &    & Gradient descent & &
			\\
			\hline
			\cite{Ginige2025} & MU-MISO  &   &   &   &  \checkmark  & Gradient descent-ascent & & \checkmark
			\\
			\hline
			\cite{Sun2025} & MU-MIMO  &   &   & \checkmark  &    & MM & \checkmark &
			\\
			\hline
			\cite{Papazafeiropoulos2025} & MU-MIMO  &   &   & \checkmark  &    & AO+PGA & &
			\\
			\hline
			\cite{Hu2024} & CF-MIMO &   &   & \checkmark  &    & AO &  &
			\\
			\hline
			\cite{Li2024} & UL CF-MIMO &   &   & \checkmark  &    & AO & \checkmark &
			\\
			\hline
			\cite{Papazafeiropoulos2024} & P2P MIMO &   &   & \checkmark  &    & PGA &  &
			\\
			\hline
			\cite{Bahingayi2025} & P2P MIMO &   &   & \checkmark  &    & AO &  &
			\\
			\hline
			\textbf{This work} & MU-MISO &  \checkmark &  \checkmark & \checkmark  &    & Consensus-ADMM+AO  & \checkmark & \checkmark
			\\
			\hline
	\end{tabular}}
\end{table*}

{Inspired by \cite{An2023a}, SIM-enabled wireless communication systems represent an innovative concept, with a growing body of studies emerging in the current literature.  In \cite{An2023} and \cite{An2025}, based on projected gradient ascent (PGA) methods, the joint optimization algorithms of transmit power allocation at base station (BS) and wave-field beamforming at SIM  were proposed to maximize the sum-rate (SR) in SIM-assisted multiuser multiple-input single-output (MU-MISO) downlink  systems.  With statistical statistical channel state information (CSI), the optimal beamforming was addressed to maximize the SR by \cite{Papazafeiropoulos2024a} in SIM-assisted MU-MISO systems. In \cite{Lin2024}, a novel low-earth orbit satellite MU-MISO communication system assisted by SIM was introduced to maximize the ergodic SR with statistical CSI. In \cite{Liu2025}, a deep reinforcement learning (DRL)-based optimization algorithm was designed to maximize the SR in MU-MISO downlink systems.  In \cite{Kavianinia2025}, a methodology was proposed to maximize the sum secrecy rate in  SIM-assisted MU-MISO downlink systems. In integrated sensing and communication (ISAC) systems, the optimal beamforming was studied to maximize the SR in \cite{Li2024a}. In \cite{Ginige2025}, the minimum signal-to-interference-plusnoise ratio (MSINR) maximization problem was studied for MU-MISO downlink systems under both statistical and instantaneous CSI conditions. Besides, in \cite{Sun2025}, a dual-polarized stacked metasurface transceiver architecture was proposed in  MU-MIMO  systems to solve the generalized SR maximization problem. Besides, in \cite{Sun2025}, a dual-polarized stacked metasurface transceiver architecture was proposed for MU-MIMO systems, where the generalized sum-rate maximization problem was addressed using a majorization-minimization (MM)-based algorithm. In \cite{Papazafeiropoulos2025}, a joint optimization algorithm of BS-side and channel-side SIM phase shifts was proposed to maximize the sum spectral efficiency for double-SIM-assisted MU-MIMO systems. In \cite{Hu2024}, a joint optimization algorithm was proposed to maximize the SR in SIM-assisted cell-free MIMO (CF-MIMO) systems. \cite{Li2024}  proposed an SIM-based architecture for uplink (UL) CF-MIMO systems to enhance achievable spectral and energy efficiency (EE).  Furthermore,  in \cite{Papazafeiropoulos2024}, a beamforming optimization algorithm based on PGA methods was proposed to maximize the achievable rate in SIM-assisted Point-to-Point (P2P) MIMO systems. In \cite{Bahingayi2025}, two optimization problems: achievable rate maximization and inter-stream interference minimization were studied in SIM-assisted P2P MIMO systems. }

\subsection{Motivation and Contributions}
Current research on multiuser systems predominantly focuses on the SR maximization problem, e.g., \cite{An2023,Liu2025,Kavianinia2025,Hu2024,Sun2025,Li2024a,Lin2024,Papazafeiropoulos2025,An2025}. Although SR maximization ensures high overall throughput,  it may lead to potential rate fairness issues, as users with poor channel conditions might be assigned limited or even no resources (e.g., transmit power), thereby resulting in low or zero achievable rates. This can negatively affect the overall user experience and quality of service (QoS). Although SIM has shown the substantial potential of mitigating multiuser interference~\cite{An2023a}, existing works have largely overlooked how SIM can be leveraged to improve rate fairness in multiuser systems.  A recent study~\cite{Ginige2025} made a valuable step in this direction by addressing rate fairness through MSINR maximization. While insightful, MSINR maximization can be viewed as a simplified surrogate of minimum rate (MR) maximization. Given that the logarithmic function in the rate expression is inherently non-linear, the solutions to the optimization problem with the multi-user rate function as the optimization objective may not be consistent with those of the optimization problem with the multi-user SINR function as the optimization objective. 
Our work, conducted almost concurrently with~\cite{Ginige2025}, provides a more comprehensive perspective by focusing directly on rate-based fairness criteria. In particular, we investigate two commonly adopted objectives: MR maximization and geometric mean of rates (GMR) maximization, which better capture fairness at the user level and offer a more balanced trade-off between fairness and SR. In our system design, SIM is integrated with the radome of the BS and performs transmit beamforming in the EM wave domain to serve multiple users simultaneously. For clarity, a preliminary categorization and comparison of related literature, is presented in Table~\ref{tab:LiteratureComparison}. Moreover, through comparative simulation experiments, we have demonstrated that the multiuser rate performance of this work are superior to those of similar works, e.g., \cite{An2023,Ginige2025}.

Our main contributions can be summarized in detail as follows:
\begin{itemize}
\item To enhance rate fairness among users, we maximize the MR by jointly optimizing power allocation at the BS and phase shifts at the SIM. Due to the problem’s non-convexity, we propose an iterative algorithm that provides closed-form solutions. Specifically, we approximate the user rate with a tractable lower bound to reformulate the problem. By leveraging the consensus alternating direction method of multipliers (ADMM) method, we decompose the original problem into several simpler sub-problems, which allows us to tailor customized optimization methods, yielding closed-form solutions. 
\item 
To balance rate fairness and SR, we formulate the GMR maximization problem. Unlike MR maximization, it combines geometric mean function and logarithmic rate function and thus lead to strong coupling among optimization variables. By exploiting the concavity of the geometric mean function and leveraging a linear approximation, we approximately decompose it into two sub-problems: power allocation at the BS and phase shift optimization at the SIM, both of which admit closed-form solutions. Besides, this algorithm can be seamlessly applied in SR maximization problem.
\item Numerical results confirm the benefits of SIM deployment and validate the convergence of the proposed algorithms. A comparative analysis across multiple performance metrics, using SR maximization as a benchmark, reveals that MR maximization optimization ensures near-perfect fairness, as indicated by metrics such as MR, rate standard deviation, and min-rate/max-rate ratio. Meanwhile, GMR maximization achieves a balance between fairness and SR, positioning its performance between MR maximization and SR maximization. Besides, the proposed algorithms outperform existing methods from \cite{An2023} and \cite{Ginige2025} in terms of the MR and SR, respectively. Finally, our results show that SIM attains system-limit performance comparable to multi-antenna digital beamforming while reducing power consumption. 
\end{itemize}
	\begin{figure}[!t]
	\centering
	\includegraphics[width=0.47\textwidth]{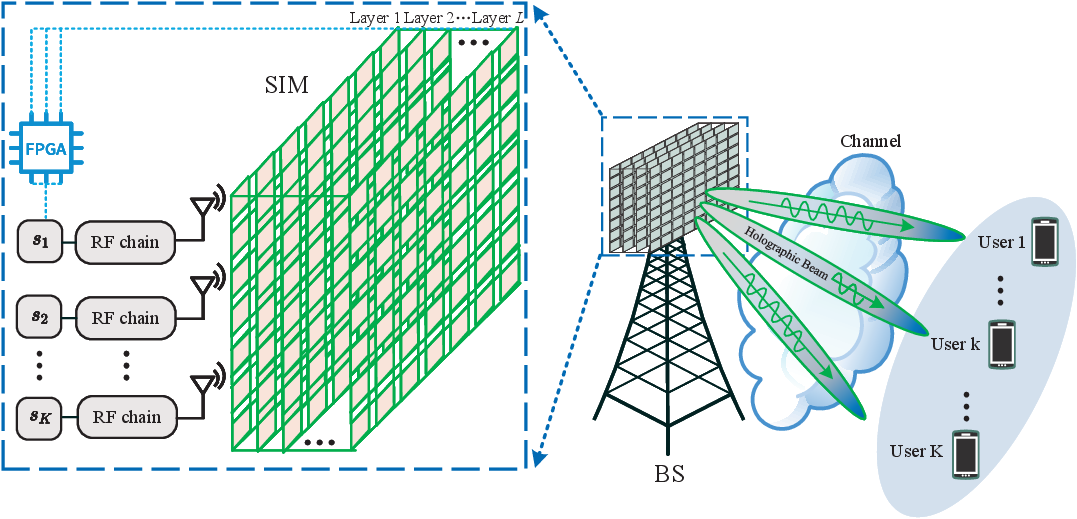}
	\caption{An SIM-assisted multiuser wireless communication system (denote layer 1, layer 2, layer $L$).}
	\label{sys_model}
\end{figure} %
\subsection{Organization and Notation}
The rest of this paper is organized as follows. Section II presents system model and problem formulation. The proposed optimization algorithms for the MR maximization optimization and GMR maximization are shown in Sections III and IV, respectively. Section V evaluates the performance of the proposed algorithms. Finally, Section VI concludes the paper.
	
\textit{Notation}: Bold upper-case and lower-case letters represent matrices and vectors, respectively, while italic letters denote scalars. $( \cdot )^{T}$, $(  \cdot  )^{\rm{*}}$, and $\left(  \cdot  \right)^{H}$ represent transpose, conjugate, and conjugate transpose, respectively. $\odot$ denotes the Hadamard product. The real part of $ \left( \cdot  \right)$ is denoted by $\Re \left( \cdot  \right)$.  $\arg\left(  \cdot  \right)$ denotes the phase information of $\left(  \cdot  \right)$. For a complex vector $\mathbf x$, ${\mathrm{diag}}\left(\mathbf x \right)$ represents a diagonal matrix whose diagonal entries are composed of elements of $\mathbf x$ and $\left[ \mathbf{x}\right]_{m}$ denote the $m$-th element of $\mathbf x$. For a square matrix $\mathbf{X}$,  $\mathrm{Tr}\left( \mathbf{X} \right)$, $ {\mathbf{X}}^{-1} $,  $\left| \mathbf{X} \right|$, and $\left[\mathbf{X}\right]_{a,b}$ denote its trace, inverse, determinant, and the element in the $a$-th row and $b$-th column.

\section{System Model and Problem Formulation}
\subsection{System Model}
We consider a multiuser MISO downlink wireless communication system assisted by an SIM, as shown in Fig.~\ref{sys_model}. In this system, a BS with $M$ antennas transmits signals to $K$ single-antenna users. The SIM comprises $L$ layers of stacked reconfigurable metasurfaces, which are uniformly managed by an intelligent controller, and each layer  consists of $N$ meta-atoms. To avoid performance degradation caused by two-hop path loss and reduce the transmission burden of real-time information exchanges, the SIM is integrated with the antenna radome of the BS\cite{An2025}. Additionally, to prevent interference from undesired diffraction, scattering, and ambient noise, the components are encapsulated within a sealed vessel\cite{An2023a}. Specifically, the EM wave emitted by the BS passes through each layer of the SIM in sequence, with each layer using a controller to adjust the characteristics of the incident EM wave, such as its phase and/or amplitude, to precisely manipulate the wavefront and achieve multiuser precoding automatically for enhancing overall transmission  performance~\cite{An2024}.

We define the set of transmit antennas, SIM layers, all elements on each metasurface layer, and users by $\mathcal{M}  = \left\{ { 1,2,...,M} \right\}$,  $\mathcal{L}  = \left\{ { 1,2,...,L} \right\}$, $\mathcal{N}  = \left\{ { 1,2,...,N} \right\}$, and $\mathcal{K}  = \left\{ { 1,2,...,K} \right\}$, respectively. Taking into account the balance between energy efficiency and implementation costs, we solely focus on the phase-shifting capability of these elements, as \cite{Kavianinia2025,Liu2025,Hu2024,Sun2025,Li2024a,Lin2024,Papazafeiropoulos2025,An2025}. Hence, we assume that the amplitude coefficients of all elements are set to unit. Then, we define the coefficient vector of the $l$-th layer of the SIM  as  $e^{j \boldsymbol{\theta}_{l}}=\left[e^{j \theta_{l,1}}, e^{j \theta_{l,2}}, \cdots, e^{j \theta_{l,N}} \right]$, where $\theta_{l,n} \in[0, 2\pi)$ is the phase shift of the $n$-th meta-atom on the $l$-th metasurface layer, for $n \in \mathcal{N}$ and $l \in \mathcal{L}$. And then we can define   $\boldsymbol{\Theta}_{l} = \mathrm{diag}\left(e^{j \boldsymbol{\theta}_{l}} \right)$. 
Let $\mathbf{W}_{l} \in \mathbb{C}^{N \times N}$, for $l \in \mathcal{L}$, $l \neq 1$, and  $\mathbf{W}_{1}=\left[\mathbf{w}_{1,1}, \mathbf{w}_{1,2}, \cdots, \mathbf{w}_{1,M} \right] \in \mathbb{C}^{N \times M}$   denote the channel response matrix from the $\left( l-1\right) $-th metasurface layer to the $l$-th metasurface layer, and the channel response matrix from the BS to the 1-th metasurface layer, respectively.  For $k \in \mathcal{K}$,  $\mathbf{h}_{\text{SIM-}k}^H \in \mathbb{C}^{1 \times N}$ stands for the channel vector from the SIM to the $k$-th user. According to the Rayleigh-Sommerfeld diffraction theory, the element of the channel response matrix $\mathbf{W}_{l}$ in the $n$-th row and $n'$-th column, for $n,n' \in \mathcal{N}$, $l \in \mathcal{L}$, $l \neq 1$, is given by \cite{An2023,An2023a,An2024,An2024a}
\begin{equation}\label{eq_ChannelWl}
	\left[\mathbf{W}_{l}\right]_{n,n'} =\frac{d_l d_w  d_t}{(L-1) d_{n,n'}^2}\left(\frac{1}{2 \pi d_{n,n'}}- \frac{j}{\lambda_c}\right)e^{\frac{j 2 \pi d_{n,n'}}{\lambda_c}},
\end{equation}
where $d_l$ and $d_w$ represent the length and width of the SIM meta-atoms, $d_t$ denotes the thickness of the whole SIM, $d_{n,n'}$ is the distance between the $n$-th meta-atom on the $l$-th layer and the $n'$-th meta-atom on the $(l-1)$-th layer, and $\lambda_c$ denotes the carrier wavelength. Since the SIM is mounted on the BS, the transmission from BS to SIM  falls within the near-field line-of-sight (LoS) regime. Based on the spherical wave propagation characteristics, the meta-atom of the channel response matrix $\mathbf{W}_{1}$ in the $n$-th row and $m$-th column, for $n \in \mathcal{N}$, $m \in \mathcal{M}$, is given by\cite{An2023,An2023a,An2024,An2024a}
\begin{equation}\label{eq_ChannelW1}
	\left[\mathbf{W}_{1}\right]_{n,m} =\frac{\lambda_c}{4 \pi d_{n,m}}e^{\frac{-j 2 \pi d_{n,m}}{\lambda_c}},
\end{equation}
where $d_{n,m}$ is the distance between the $n$-th meta-atom on the 1st layer and the $m$-th antenna of the BS. Then, the equivalent EM wave domain beamforming matrix of the SIM can be given by
\begin{equation}\label{eq_SIM_beamforming}
	\mathbf{G}=\boldsymbol{\Theta}_{L} \mathbf{W}_{L} \boldsymbol{\Theta}_{L-1} \mathbf{W}_{L-1} \cdots \boldsymbol{\Theta}_{2} \mathbf{W}_{2} \boldsymbol{\Theta}_{1} \in \mathbb{C}^{N \times N } .
\end{equation}

To focus solely on the wave-based beamforming gain enabled by the SIM and avoid the complications of antenna selection, we assume $K = M$, as in \cite{An2023,An2023a,An2024,Lin2024,Papazafeiropoulos2024,Li2024,An2024a}. This simplification removes the need to pre-select $K$ transmit antennas from $M$ candidates, which is a typical challenge in conventional MIMO systems with limited RF chains. The joint design of antenna selection and SIM-based beamforming is left for future work. Specifically, each antenna at the BS just transmits a signal intended for one user, while the BS focuses solely on power allocation, without participating in digital beamforming design. Additionally, we assume knowledge of the instantaneous CSI for all channels, which can be obtained by  leveraging advanced channel estimation methods proposed in \cite{Yao2024,Nadeem2024}. It is worth noting that the algorithms presented in this paper can be easily adapted, with minor modifications, to scenarios where both the BS and SIM participate in beamforming design. 

Consequently, for $k \in \mathcal{K}$, the received signal at the $k$-th user can be expressed as
	\begin{equation}\label{eq_received signal}
			y_{k} = \mathbf{h}_{\text{SIM-}k}^H \mathbf{G} \mathbf{W}_{1}  \mathbf{x} + n_{k}, 
	\end{equation}
where $\mathbf{x}=\left[\sqrt{p_1} s_1, \sqrt{p_2} s_2 \cdots, \sqrt{p_K} s_K \right] \in \mathbb{C}^{M \times 1 }$, $p_k$ denotes the BS transmit power allocated to the $k$-th user,  $s_k$ is the desired signal for the $k$-th user and follows the Gaussian distribution with zero mean and unit variance, i.e., $s_{k}\sim\mathcal{CN} (0,1)$, and $n_{k}\sim\mathcal{CN} (0,\sigma _{k}^2)$ is the independent and identically distributed (i.i.d.) Gaussian noise. The received SINR at the $k$-th user is given by
\begin{equation}
	\mathrm{SINR}_k =  \frac{\left|  \mathbf{h}_{\text{SIM-}k}^H \mathbf{G} \mathbf{w}_{1,k}\right|^2  p_k}{\sum\limits_{m=1, m\ne k}^{K} \left|  \mathbf{h}_{\text{SIM-}k}^H \mathbf{G} \mathbf{w}_{1,m} \right|^2 p_m + \sigma _{k}^2} .
\end{equation}
Then, the achievable rate of the $k$-th user can be expressed by
\begin{equation}\label{eq_rate}
	r_k =  \log_2 \left(   1 + \mathrm{SINR}_k  \right).
\end{equation}
It is worth highlighting that existing research predominantly emphasizes SR maximization, typically compromising user fairness, which underscores the motivation for our study.
\subsection{Problem Formulation}
In this work, we focus on two commonly encountered rate fairness optimization problems: the MR maximization and the GMR maximization, to equitably evaluate the effect of SIM on  enhancing rate fairness.
\subsubsection{MR Maximization}
The goal here is to maximize the minimum achievable rate across all $K$ users, i.e., $\max\{\min{r_k}\}$ for $k\in \mathcal{K}$, by jointly optimizing the power allocation coefficients $\{p_k\}$
at the BS and the beamforming matrices $\{\boldsymbol{\Theta}_{l}\}$ for $l\in \mathcal{L}$
at the SIM, with the total power constraint at the BS and the transmission coefficient constraints at the SIM. 
To facilitate the optimization problem, we define the auxiliary variable $\rho_k = \sqrt{p_k}$ for $k \in \mathcal{K}$. Mathematically, the MR maximization problem can be expressed as
\begin{align} \label{P1} 
	\mathrm{\left(P 1\right)}: &\mathop {\max }\limits_{ \{\rho_k,\boldsymbol{\Theta}_{l}\},\gamma} \mathop  \gamma   
	\\ \mbox{s.t.}\quad
	&r_k \ge  \gamma ,k \in \mathcal{K}, \tag{\ref{P1}{a}} \label{P1a}
	\\
	& \sum_{k=1}^K \rho_k^2 \le P_{\max} \tag{\ref{P1}{b}} \label{P1b},
	\\
	& \rho_k \ge  0  ,k \in \mathcal{K} \tag{\ref{P1}{c}} \label{P1c},
	\\
	&0 \le \theta_{l,n} \le 2\pi ,n \in \mathcal{N}, l \in \mathcal{L}. \tag{\ref{P1}{d}} \label{P1d}
\end{align}
where $P_{\max}$ is the maximum allowable total transmit power at the BS. From (P1), we can see that rates of all users are always larger than $\gamma$, which represents the required minimum QoS for all users. Thus, the rates of the users experiencing poor channel conditions are also guaranteed, rather than being forced to be few or even zero in SR optimization. As a result, in the MR maximization  framework, the rate fairness of all users is improved to a certain extent. 
\subsubsection{GMR Maximization}
Clearly,  MR maximization may not fully leverage the system's potential for maximizing overall SR. In contrast, the second problem, GMR maximization, aims to balance individual users' rates by promoting a more equitable distribution, potentially improving overall system performance without disproportionately benefiting any single user. With the same constraints at the BS and SIM, the GMR maximization problem can be formulated as
\begin{align} \label{P2} 
	\mathrm{\left(P 2\right)}: &\mathop {\max }\limits_{ \{\rho_k,\boldsymbol{\Theta}_{l}\}} \mathop     {\left( \prod \limits_{k=1}^K r_k \right)^{1/K}}
	\\ \mbox{s.t.}\quad
	&\eqref{P1b}, \eqref{P1c}, \eqref{P1d}. \tag{\ref{P2}{a}} \label{P2a}
\end{align}
Although GMR maximization may not always try to promote the rate for the most disadvantaged user, in \cite{Yu2024} it is shown that maximizing the GMR can achieve Pareto-optimal rates, therefore ensuring a degree of rate fairness along with favorable SINR. Moreover, the GMR maximization problem is computationally simpler due to the absence of the mandatory and intractable constraint \eqref{P1a} found in (P1), leading to a relatively lower algorithmic complexity. 

We can see that both (P1) and (P2) are highly non-convex, as (P1) has its non-convex constraint \eqref{P1a} and the objective function of (P2) is non-concave. 
And especially when $L$ and $N$ become very large, both problems would result in a more intricate optimization landscape, due to a high prevalence of local optima and saddle points in solving process. Therefore, the development of efficient algorithms is extremely crucial to advance the practical applications of SIM. Next, we separately design solving algorithms for above two problems.

\section{Solving Algorithm for MR Maximization}
\subsection{MR Maximization Algorithm Design}
The existence of the auxiliary variable $\gamma$ in \eqref{P1a} forces us to directly deal with $r_k$ in order to avoid the intricate coupling between  $\{\rho_k,\boldsymbol{\Theta}_{l}\}$ and $\gamma$. Therefore, given a feasible solution, we intend to develop a feasible lower bound for $r_k$ to construct the approximate optimization problem of (P1)\cite{Naghsh2019,Fang2023,Fang2023a}. 

It is noted that $r_k$ is not jointly
concave with respect to (w.r.t.) $\{\rho_k\}$ and $\{\boldsymbol{\Theta}_{l}\}$. For convenience, i.e., to avoid dealing with the intricate coupling between $\{\rho_k\}$ and $\{\boldsymbol{\Theta}_{l}\}$ in $r_k$ of \eqref{P1a}, we first integrate $\{\rho_k\}$ and $\{\boldsymbol{\Theta}_{l}\}$ into an auxiliary variable $\{\mathbf{A}_{k}\}$. According to the partitioned  matrix inversion lemma, $r_k$ in \eqref{eq_rate} is equivalent to 
\begin{equation}\label{eq_re_rate}
	r_k=\log_2 \left( \mathbf{p}  \mathbf{A}_{k}^{-1} \mathbf{p}^T\right),
\end{equation} 	 
where
\begin{equation*}\label{eq_Ak}
	\begin{aligned}
		&\mathbf{p}=[1,0],
		\\
		&\mathbf{A}_{k}\!\!=\!\!\begin{bmatrix} 1 & (  \mathbf{h}_{\text{SIM-}k}^H \mathbf{G} \mathbf{w}_{1,k}  \rho_k)^H \\   \mathbf{h}_{\text{SIM-}k}^H \mathbf{G} \mathbf{w}_{1,k}  \rho_k & \sum\limits_{m=1}^{K} \left|  \mathbf{h}_{\text{SIM-}k}^H \mathbf{G} \mathbf{w}_{1,m} \rho_m \right|^2  \!\!\!+\!\! \sigma _{k}^2 \end{bmatrix} \in \mathbb{C}^{2 \times 2}.
	\end{aligned}
\end{equation*} 
We can find that $r_k$ in \eqref{eq_re_rate} is convex w.r.t. $\mathbf{A}_{k}$. Accordingly, due to the first-order property of convex function, we can derive the lower bound of $r_k$ given a certain feasible solution. Assuming that $\{\rho_k^{\left(i \right)}\}$ and $\{\boldsymbol{\Theta}_{l}^{\left(i \right)}\}$ are arbitrary feasible solutions obtained in the $i$-th round, we can then obtain the corresponding  $\mathbf{G}^{\left(i \right)}$, ${\mathbf{A}}_{k}^{\left(i \right)}$, and $\mathrm{SINR}_k^{\left(i \right)}$. Consequently, the lower bound of $r_k$ in the $(i+1)$-th round can be expressed as
\begin{align}\label{eq_r_k_hat}
\hat{r}_{k,i} \triangleq &\frac{1}{\ln2}
\left(	c_k^{\left(i \right)} -2 \Re \left( b_{k,12}^{\left(i \right)} \mathbf{h}_{\text{SIM-}k}^H \mathbf{G} \mathbf{w}_k^{1}  \rho_k \right) \right.  \nonumber
\\
&\left. - b_{k,22}^{\left(i \right)} \sum\limits_{m=1}^{K} \left|  \mathbf{h}_{\text{SIM-}k}^H \mathbf{G} \mathbf{w}_m^{1} \rho_m \right|^2 \right) \le r_k,
\end{align}
where 
\begin{align}
	&c_k^{\left(i \right)}=\ln\left(\mathbf{p} ({\mathbf{A}}_{k}^{\left(i \right)})^{-1} \mathbf{p}^T \right)+ \mathrm{Tr}\left({\mathbf{B}}_{k}^{\left(i \right)} {\mathbf{A}}_{k}^{\left(i \right)}\right)-b_{k,11}^{\left(i \right)}-\sigma_k^2 b_{k,22}^{\left(i \right)}, \nonumber
	\\
	&{\mathbf{B}}_{k}^{\left(i \right)} = ({\mathbf{A}}_{k}^{\left(i \right)})^{-1} \mathbf{p}^T \left( \mathbf{p}  ({\mathbf{A}}_{k}^{\left(i \right)})^{-1} \mathbf{p}^T \right)^{-1} \mathbf{p} ({\mathbf{A}}_{k}^{\left(i \right)})^{-1},\nonumber
\end{align}
and $b_{k,uv}^{\left(i \right)}$ for $u,v\in\{1,2\}$ are sub-blocks of ${\mathbf{B}}_{k}^{\left(i \right)}$, i.e., 
\begin{align}
\mathbf{B}_{k}^{\left(i \right)}=\begin{bmatrix}  b_{k,11}^{\left(i \right)}  &  b_{k,12}^{\left(i \right)}  \\   b_{k,21}^{\left(i \right)}   &   b_{k,22}^{\left(i \right)}  \end{bmatrix}. \nonumber
\end{align}
In particular, we have
\begin{align}
	&b_{k,11}^{\left(i \right)}=1+ \mathrm{SINR}_k^{\left(i \right)},\label{eq_b11}
	\\
	& b_{k,12}^{\left(i \right)}=\left(b_{k,21}^{\left(i \right)}\right)^*= -\mathrm{SINR}_k^{\left(i \right)}/\mathbf{h}_{\text{SIM-}k}^H \mathbf{G}^{\left(i \right)} \mathbf{w}_{1,k}  \rho_k^{\left(i \right)} \label{eq_b12},
	\\
	&b_{k,22}^{\left(i \right)}= b_{k,12}^{\left(i \right)} b_{k,21}^{\left(i \right)} / b_{k,11}^{\left(i \right)}. \label{eq_b22}
\end{align}

Then, given $\{\rho_k^{\left(i \right)}\}$ and $\{\boldsymbol{\Theta}_{l}^{\left(i \right)}\}$, we can get the next feasible solutions $\{\rho_k^{\left(i+1 \right)}\}$ and $\{\boldsymbol{\Theta}_{l}^{\left(i+1 \right)}\}$ by solving 
\begin{align} \label{P1.1} 
	\mathrm{\left(P1.1\right)}: &\{\rho_k^{\left(i+1 \right)}, \boldsymbol{\Theta}_{l}^{\left(i+1 \right)} \}= \mathop {\arg \max }\limits_{ \{\rho_k,\boldsymbol{\Theta}_{l}\},\gamma} \mathop  \gamma   
	\\ \mbox{s.t.}\quad
	&\hat{r}_{k,i} \ge  \gamma ,k \in \mathcal{K}, \eqref{P1b}, \eqref{P1c}, \eqref{P1d}. \tag{\ref{P1.1}{a}} \label{P1.1a} 
\end{align}
Unfortunately, the optimization problem (P1.1) still cannot be solved directly due to the coupling between $\{\rho_k\}$ and $\{\boldsymbol{\Theta}_{l}\}$ in the expression of $\hat{r}_{k,i}$, e.g., \eqref{eq_r_k_hat}. Since consensus-ADMM excels in solving large-scale and distributed optimization problems by decomposing an original problem into smaller, more manageable sub-problems that can be solved in parallel \cite{Huang2016,Fang2024}, we develop a consensus-ADMM-based algorithm to solve (P1.1). One can observe that $\gamma$, $\{\rho_k\}$, and $\{\boldsymbol{\Theta}_{l}\}$ appear in all $K$ rate constraints in \eqref{P1.1a}, so we have to introduce auxiliary variables $z_{km}=\mathbf{h}_{\text{SIM-}k}^H \mathbf{G} \mathbf{w}_{1,m} \rho_m $ and $\eta_k=\gamma$ for $k,m \in \mathcal{K}$ to transform (P1.1) into a decomposable structure. Then, (P1.1) can be equivalently given by
	\begin{align} \label{P3} 
	\mathrm{\left(P1.2\right)}: &\mathop {\max }\limits_{ \{z_{km},\eta_k,\rho_k,\boldsymbol{\Theta}_{l}\},\gamma} \mathop  \gamma   
	\\ \mbox{s.t.}\quad
	&c_k^{(i)} - 2 \Re \left( b_{k,12}^{(i)} z_{kk}\right)-  \sum\limits_{m=1}^{K} b_{k,22}^{(i)} z_{km} z_{km}^*  \nonumber
	\\
	&  \ge   \ln2 \eta_k ,k \in \mathcal{K}, \tag{\ref{P3}{a}} \label{P3a}
	\\
	&z_{km}=\mathbf{h}_{\text{SIM-}k}^H \mathbf{G} \mathbf{w}_{1,m} \rho_m , k,m \in \mathcal{K},\tag{\ref{P3}{b}} \label{P3b}
	\\
	&\eta_k = \gamma ,k \in \mathcal{K}, \tag{\ref{P3}{c}} \label{P3c}
     \\
     & \eqref{P1b}, \eqref{P1c}, \eqref{P1d}.	\tag{\ref{P3}{d}} \label{P3d}
\end{align}

Let $\mathscr{F}_{z,\eta}$, $\mathscr{F}_{\rho}$, and $\mathscr{F}_{\theta}$  represent the feasible sets of variables $\{z_{km},\eta_k\}$,  $\{\rho_k\}$, and $\{\boldsymbol{\Theta}_{l}\}$ satisfying the inequality constraints \eqref{P3a}, \eqref{P1b}, \eqref{P1c}, and \eqref{P1d}, respectively. Then, the indicator functions corresponding to $\mathscr{F}_{z,\eta}$, $\mathscr{F}_{\rho}$, and $\mathscr{F}_{\theta}$ can be defined by
	\begin{equation}\label{eq_indicator}
	\begin{aligned}
		\mathbb{I}_{\mathscr{F}_{z,\eta}}\left(\{z_{km},\eta_k\} \right) &=
		\begin{cases}
			0, \quad \text{if} \; \{z_{km},\eta_k\} \in \mathscr{F}_{z,\eta},
			\\ 
			-\infty, \quad \text{otherwise}.
		\end{cases}
		\\
		\mathbb{I}_{\mathscr{F}_{\rho}}\left(\{\rho_k\} \right) &=
		\begin{cases}
			0, \quad \text{if} \; \{\rho_k\} \in \mathscr{F}_{\rho},
			\\ 
			-\infty, \quad \text{otherwise}.
		\end{cases} 
		\\
		\mathbb{I}_{\mathscr{F}_{\theta}}\left(\{\boldsymbol{\Theta}_{l}\} \right) &=
		\begin{cases}
			0, \quad \text{if} \; \{\boldsymbol{\Theta}_{l}\} \in \mathscr{F}_{\theta},
			\\ 
			-\infty, \quad \text{otherwise}.
		\end{cases}
	\end{aligned}
\end{equation}
Integrating the inequality constraints into the objective function via \eqref{eq_indicator}, (P1.2) can be equivalently transformed to the following consensus equality-constrained problem
\begin{align} \label{P1.3}
	\mathrm{\left(P 1.3 \right)}: &\mathop {\max }\limits_{ \{z_{km},\eta_k,\rho_k,\boldsymbol{\Theta}_{l}\},\gamma}   \mathbb{I}_{\mathscr{F}_{z,\eta}}\left(\{z_{km},\eta_k\} \right) + \mathbb{I}_{\mathscr{F}_{\rho}}\left(\{\rho_k\} \right)  \nonumber
	\\
	&+\mathbb{I}_{\mathscr{F}_{\theta}}\left(\{\boldsymbol{\Theta}_{l}\} \right)+ \gamma \\ \mbox{s.t.}\quad
	& \eqref{P3b}, \eqref{P3c}. \label{P3'} \tag{\ref{P1.3}{a}}
\end{align}
According to coupling relationship introduced by constraints,  the optimization variables can be split into four separate variable blocks, i.e., $\{z_{km},\eta_k\}$, $\{\rho_k\}$, $\{\boldsymbol{\Theta}_{l}\}$, and $\gamma$. 
Therefore, the augmented Lagrangian of (P1.3) is given by
\begin{align} \label{eq_Lagrangian} 
	&{\cal L}_{P1.3}\left( \{z_{km},\eta_k,\rho_k,\boldsymbol{\Theta}_{l},z_{d,km},\eta_{d,k}\},\gamma  \right) = \nonumber
	\\
	&\mathbb{I}_{\mathscr{F}_{z,\eta}}\left(\{z_{km},\eta_k\} \right) + \mathbb{I}_{\mathscr{F}_{\rho}}\left(\{\rho_k\} \right)+\mathbb{I}_{\mathscr{F}_{\theta}}\left(\{\boldsymbol{\Theta}_{l}\} \right)+ \gamma - \nonumber
	\\
	&\frac{\rho}{2} \sum_{k=1}^K \sum_{m=1}^K \left| z_{km}-\mathbf{h}_{\text{SIM-}k}^H \mathbf{G} \mathbf{w}_{1,m} \rho_m + z_{d,km}\right|^2 - \nonumber
	\\
	&   \frac{\rho}{2} \sum_{k=1}^K \left( \eta_k-\gamma+ \eta_{d,k}  \right)^2,
\end{align}
where $\rho>0$ denotes the penalty parameter for step-size control,  $z_{d,km}$ and $\eta_{d,k}$, for $k,m \in \mathcal{K}$, are scaled dual variables associated with  constraint \eqref{P3b} and \eqref{P3c}. That is to say now we can maximize ${\cal L}_{P1.3}\left( \{z_{km},\eta_k,\rho_k,\boldsymbol{\Theta}_{l},z_{d,km},\eta_{d,k}\},\gamma  \right)$ w.r.t. $\{z_{km},\eta_k\}$, $\{\rho_k\}$, $\{\boldsymbol{\Theta}_{l}\}$, and $\gamma$ to find the solution of (P1.3). 

To solve (P1.3), we need to alternately update the four variable blocks and the dual variables $\{z_{d,km},\eta_{d,k}\}$ in each round of iteration. Let $\{z_{km}^{\left(j \right)},\eta_k^{\left(j \right)}\}$, $\{\rho_k^{\left(j \right)}\}$, $\{\boldsymbol{\Theta}_{l}^{\left(j \right)}\}$, $\gamma^{\left(j \right)}$, and $\{z_{d,km}^{\left(j \right)},\eta_{d,k}^{\left(j \right)}\}$ be the solutions after the $j$-th consensus-ADMM iteration procedure of (P1.3). Assuming that $j^*$ is the number of consensus-ADMM iterations required to solve (P1.3), we have $\rho_k^{\left(i+1 \right)}=\rho_k^{\left(j^* \right)}$, $\boldsymbol{\Theta}_{l}^{\left(i+1 \right)}=\boldsymbol{\Theta}_{l}^{\left(j^* \right)}$, for $k \in \mathcal{K}, l \in \mathcal{L}$, which can be used to generate $\hat{r}_{k,i+1}$ and then update optimization problem (P1.2). To summarize, the whole algorithm consists of two layers of iterations, the outer iteration is used to update (P1.2) and the inner consensus-ADMM iteration is used to solve (P1.2). We next show the details of the five steps in  the $(j+1)$-th inner consensus-ADMM iteration. 

\textbf{STEP 1}: Updating $\{z_{km}^{\left(j+1 \right)},\eta_k^{\left(j+1 \right)}\}$ \\
From \eqref{eq_Lagrangian}, given $\{z_{km}^{\left(j \right)},\eta_k^{\left(j \right)}\}$, $\{\rho_k^{\left(j \right)}\}$, $\{\boldsymbol{\Theta}_{l}^{\left(j \right)}\}$, $\gamma^{\left(j \right)}$, and $\{z_{d,km}^{\left(j \right)},\eta_{d,k}^{\left(j \right)}\}$, we can update $\{z_{km},\eta_k\}$ by solving   
\begin{align}
	\{z_{km}^{\left(j+1 \right)},\eta_k^{\left(j+1 \right)}\}=& \mathop{\arg\max}\limits_{\{z_{km},\eta_k\}}  {\cal L}_{P1.3}\left( {\{z_{km},\eta_k,\rho_k^{\left(j \right)},}\right.  \nonumber
	\\
	&\left. {\boldsymbol{\Theta}_{l}^{\left(j \right)},z_{d,km}^{\left(j \right)},\eta_{d,k}^{\left(j \right)}\},\gamma^{\left(j \right)}}  \right).\label{eq_zetaADMM} 
\end{align}
By omitting the irrelevant terms $\mathbb{I}_{\mathscr{F}_{\rho}}\left(\{\rho_k^{\left(j \right)}\} \right)$, $\mathbb{I}_{\mathscr{F}_{\theta}}\left(\{\boldsymbol{\Theta}_{l}^{\left(j \right)}\} \right)$, and $ \gamma^{\left(j \right)}$ in \eqref{eq_Lagrangian}, \eqref{eq_zetaADMM} is equivalent to
\begin{align} 
	\mathrm{\left(P 1.3.1 \right)}\!\!:\!&\mathop {\min  }\limits_{ \{z_{km},\eta_k\}  }  \sum_{k=1}^K\! \left(  {\sum_{m=1}^K  \left| z_{km} \!\!\!-\!\!\mathbf{h}_{\text{SIM-}k}^H \mathbf{G}^{\left(j \right)} \mathbf{w}_{1,m} \rho_m^{\left(j \right)} \!+\!\! z_{d,km}^{\left(j \right)}\right|^2}  \nonumber
	\right.\\
    & \left. {+ \left( \eta_k-\gamma^{\left(j \right)}+ \eta_{d,k}^{\left(j \right)}  \right)^2} \right)
	\quad \mbox{s.t.}\quad
	\eqref{P3a}. \label{P4.1} 
\end{align}
It can be found that for $ k,m \in \mathcal{K}$, $z_{km}$ and $\eta_k$ are decoupled in both objective function and constraint \eqref{P3a}. Therefore, (P1.3.1) can be decomposed into $K$ convex independent quadratically constrained quadratic programming with only one constraint (QCQP-1), which can be directly solved in closed-form. Define the dual variable associated with the $k$-th constraint of \eqref{P3a} by $\lambda_k \ge 0$. The Lagrangian w.r.t. $\{z_{km}\},\eta_k,\lambda_k$, for $ m \in \mathcal{K}$, of the $k$-th sub-problem is given by 
\begin{align} 
	&{\cal L}_{k,P1.3.1}\left(\{z_{km}\},\eta_k,\lambda_k \right) = \nonumber
	\\
	&\sum_{m=1}^K  \left| z_{km} \!-\! \mathbf{h}_{\text{SIM-}k}^H \mathbf{G}^{\left(j \right)} \mathbf{w}_{1,m} \rho_m^{\left(j \right)} \!+\!\! z_{d,km}^{\left(j \right)}\right|^2  \!+\! \left(\! \eta_k \!\!- \! \gamma^{\left(j \right)} \!\!+\! \eta_{d,k}^{\left(j \right)}  \right)^2 \nonumber
	\\
	&  \!+\! \lambda_k \!\! \left( \!\! \ln2\eta_k \!+\! 2 \Re \left( b_{k,12}^{(i)} z_{kk}\right) \!\!+\!\!   \sum\limits_{m=1}^{K} b_{k,22}^{(i)} z_{km} z_{km}^* \!\!-\! c_k^{(i)}\!\! \right)\!.\!
\end{align}
By applying the Karush-Kuhn-Tucker (KKT) condition, we can obtain $\{z_{km}^{\left(j+1 \right)},\eta_k^{\left(j+1 \right)}\}$ as demonstrated in Table~\ref{tab1} by letting the gradient of ${\cal L}_{k,P1.3.1}\left(\{z_{km}\},\eta_k,\lambda_k \right)$ w.r.t. $z_{km}$ and $\eta_k$ be zero, where
\begin{align} \label{} 
	&f_{k,\lambda}=c_k^{(i)} - 2\Re\left(b_{k,12}^{(i)}\left(\mathbf{h}_{\text{SIM-}k}^H \mathbf{G}^{\left(j \right)} \mathbf{w}_{1,k} \rho_k^{\left(j \right)}-z_{d,kk}^{\left(j \right)}\right)\right)-\nonumber
	\\
	& b_{k,22}^{(i)}\sum\limits_{m=1}^{K} \left| \mathbf{h}_{\text{SIM-}k}^H \mathbf{G}^{\left(j \right)} \mathbf{w}_{1,m} \rho_m^{\left(j \right)}\!-\!z_{d,km}^{\left(j \right)} \right|^2\! \!\!-\!\! \ln2 \left(\gamma^{\left(j \right)} -\eta_{d,k}^{\left(j \right)}\right). \nonumber
	\end{align} 
With the complementary slackness condition, there is $\lambda_k=0$ if $f_{k,\lambda} \ge 0$. While, if $f_{k,\lambda} < 0$,  $z_{km}^{\left(j+1 \right)}$ and $\eta_k^{\left(j+1 \right)}$ are both monotonically decreasing w.r.t. $\lambda_k$. In this case, $\lambda_k$ can be effectively found by bisection searching.

\begin{table}[!t]
	\normalsize
	\setlength\tabcolsep{3pt}
	\renewcommand\arraystretch{1.6}
	\caption{Values of $z_{km}^{\left(j+1 \right)}$ and $\eta_k^{\left(j+1 \right)}$}
	\centering
	{
		{\begin{tabular}{|c|c|} 
				\hline
				\multirow{2}{*}{$f_{k,\lambda} \ge 0$} & $z_{km}^{\left(j+1 \right)}=\mathbf{h}_{\text{SIM-}k}^H \mathbf{G}^{\left(j \right)} \mathbf{w}_{1,m} \rho_m^{\left(j \right)}-z_{d,km}^{\left(j \right)}$ \\\cline{2-2}& $\eta_k^{\left(j+1 \right)}=\gamma^{\left(j \right)} -\eta_{d,k}^{\left(j \right)}$   
				\\
				\hline
				\multirow{3}{*}{$f_{k,\lambda} < 0$} &
				$z_{km}^{\left(j+1 \right)}\!\!=\!\!
				\begin{cases}
					\frac{\mathbf{h}_{\text{SIM-}k}^H \mathbf{G}^{\left(j \right)} \mathbf{w}_{1,k} \rho_k^{\left(j \right)}-z_{d,kk}^{\left(j \right)}-\left(\lambda_k b_{k,12}^{(i)} \right)^*}{1+\lambda_k b_{k,22}^{(i)}},\\
					\qquad \qquad\qquad \qquad \quad  \; \; \text{if} \; k=m,
					\\ 
					\frac{\mathbf{h}_{\text{SIM-}k}^H \mathbf{G}^{\left(j \right)} \mathbf{w}_{1,m} \rho_m^{\left(j \right)}-z_{d,km}^{\left(j \right)}}{1+\lambda_k b_{k,22}^{(i)}}, \;\, \text{if} \; k \ne m.
				\end{cases}$
				\\\cline{2-2}& $\eta_k^{\left(j+1 \right)}=\gamma^{\left(j \right)} -\eta_{d,k}^{\left(j \right)}-  \ln 2 \lambda_k/2$
				\\
				\hline
	\end{tabular}}}
	\label{tab1} 
\end{table}	

\textbf{STEP 2}: Updating $\{\rho_k^{\left(j+1\right)}\}$: \\
Given $\{z_{km}^{\left(j+1 \right)},\eta_k^{\left(j+1 \right)}\}$, $\{\rho_k^{\left(j \right)}\}$, $\{\boldsymbol{\Theta}_{l}^{\left(j \right)}\}$, $\gamma^{\left(j \right)}$, and $\{z_{d,km}^{\left(j \right)},\eta_{d,k}^{\left(j \right)}\}$, similarly we can update $\{\rho_k\}$ by solving 
\begin{align}
	\{\rho_k^{\left(j+1\right)}\}=& \mathop{\arg\max}\limits_{\{\rho_k\}}  {\cal L}_{P1.3}\left( {\{z_{km}^{\left(j+1 \right)},\eta_k^{\left(j+1 \right)},\rho_k,}\right.  \nonumber
	\\
	&\left. {\boldsymbol{\Theta}_{l}^{\left(j \right)},z_{d,km}^{\left(j \right)},\eta_{d,k}^{\left(j \right)}\},\gamma^{\left(j \right)}}  \right),\label{eq_rhoADMM} 
\end{align}
which is equivalent to
\begin{align} \label{P3.2} 
	\mathrm{\left(P 1.3.2 \right)}:   &\mathop {\min  }\limits_{ \{\rho_k \}  } \sum_{k=1}^K  \!  \sum_{m=1}^K \left|\! z_{km}^{\left(j+1 \right)} \!\!\!\!\!-\! \mathbf{h}_{\text{SIM-}k}^H \mathbf{G}^{\left(j \right)} \mathbf{w}_{1,m} \rho_m \!\!+\!\! z_{d,km}^{\left(j \right)}\!\right|^2\! 
	\\
	\mbox{s.t.} \quad & \eqref{P1b}, \eqref{P1c}. \tag{\ref{P3.2}{a}}
\end{align}
Note that (P1.3.2) becomes a convex QCQP-1 and can be solved using the similar approach for (P1.3.1). Then, $\rho_k^{\left(j+1 \right)}$ can be given by
	\begin{align} \label{eq_rho^j+1}
		\rho_k^{\left(j+1 \right)} =
		\begin{cases}
			\max \{0,f_{k,\beta}\left(0 \right)\}, \quad \text{if} \; \sum_{k=1}^K \left(  f_{k,\beta}\left(0 \right) \right)^2 \le P_{\max},
			\\ 
			\max \{0,f_{k,\beta}\left( \hat{\beta} \right)\}, \quad \text{otherwise}, 
		\end{cases} 
	\end{align}
where 
\begin{align} 
	f_{k,\beta}\left( \beta \right) \!\!=\!\!\frac{\sum\limits_{m\!=\!1}^{K}\! \Re\!\left(\! \left(\mathbf{h}_{\text{SIM-}m}^H \mathbf{G}^{\left(j \right)} \mathbf{w}_{1,k}\right)^*\!\! \left(z_{mk}^{\left(j+1 \right)} \!\!+\!\! z_{d,mk}^{\left(j \right)}\right) \right) } {\sum\limits_{m=1}^{K} \left| \mathbf{h}_{\text{SIM-}m}^H \mathbf{G}^{\left(j \right)} \mathbf{w}_{1,k}\right|^2  \!+\! \beta}, \nonumber
\end{align}
 and $ \hat{\beta} >  0$ is the optimal dual variable associated with the transmit power constraint \eqref{P1b} when $\sum_{k=1}^K \left(  f_{k,\beta}\left(0 \right) \right)^2 > P_{\max}$. Numerical evaluations confirm that $\rho_k^{\left(j+1 \right)}$ consistently takes non-negative values in representative scenarios.


\textbf{STEP 3}: Updating $\{\boldsymbol{\Theta}_{l}^{\left(j+1 \right)}\}$: \\
Given $\{z_{km}^{\left(j+1 \right)},\eta_k^{\left(j+1 \right)}\}$, $\{\rho_k^{\left(j+1 \right)}\}$, $\{\boldsymbol{\Theta}_{l}^{\left(j \right)}\}$, $\gamma^{\left(j \right)}$, and $\{z_{d,km}^{\left(j \right)},\eta_{d,k}^{\left(j \right)}\}$, $\{\boldsymbol{\Theta}_{l}\}$ can be updated by solving
\begin{align}
	\{\boldsymbol{\Theta}_{l}^{\left(j+1 \right)}\}=& \mathop{\arg\max}\limits_{\{\boldsymbol{\Theta}_{l}\}}  {\cal L}_{P1.3}\left( {\{z_{km}^{\left(j+1 \right)},\eta_k^{\left(j+1 \right)},\rho_k^{\left(j+1 \right)},}\right.  \nonumber
	\\
	&\left. {\boldsymbol{\Theta}_{l},z_{d,km}^{\left(j \right)},\eta_{d,k}^{\left(j \right)}\},\gamma^{\left(j \right)}}  \right).\label{eq_thetaADMM} 
\end{align}
Also, \eqref{eq_thetaADMM} is equivalent to
\begin{align} \label{P3.3} 
	\mathrm{\left(P 1.3.3 \right)}\!:\!   &\mathop {\min  }\limits_{ \{{\Theta}_{l} \}  } \sum_{k=1}^K  \! \sum_{m=1}^K \left|\! z_{km}^{\left(j+1 \right)} \!\!\!\!\!-\! \mathbf{h}_{\text{SIM-}k}^H \mathbf{G} \mathbf{w}_{1,m} \rho_m^{\left(j+1 \right)} \!\!\!+\!\! z_{d,km}^{\left(j \right)}\!\right|^2\! 
	\\
	\mbox{s.t.} \quad & 0 \le \theta_{l,n} \le 2\pi ,n \in \mathcal{N}, l \in \mathcal{L}. \tag{\ref{P3.3}{a}}
\end{align}
The most significant challenge in (P1.3.3) lies in the complex coupling between phase shifts on multiple layer metasurfaces. Nevertheless, the phase shifts of all elements are completely separate in the constraint \eqref{P1d}. The fact inspires us to adopt the BCD method to update phase shifts alternately. Specifically, aiming at the phase shift of the $n$-th meta-atom on the $l$-th metasurface layer $\theta_{l,n}$, given $\{\theta_{l',m}\}$ and $\{\theta_{l,n'}\}$ for $ l' \in \mathcal{L}, m \in \mathcal{N}$ and $ n'\ne n, n' \in \mathcal{N}$, i.e., given all phase shifts except $\theta_{l,n}$, we can update $\theta_{l,n}$ by solving 
	 \begin{align} \label{P3.3ln} 
		\mathrm{\left(P 1.3.3_\emph{l,n}\right)}: &\mathop {\min}\limits_{ \theta_{l,n} } \mathop 
		-2\Re\left(t_{l,n}^{\left(j \right)} e^{j \theta_{l,n}}\right)+d_{l,n}^{\left(j \right)}
		\\
		&\mbox{s.t.} \quad 0 \le \theta_{l,n} \le 2\pi, \tag{\ref{P3.3ln}{a}} \label{P3.3lna}
	\end{align}
	where
	\begin{align}
		&t_{l,n}^{\left(j \right)}=\left[\sum_{k=1}^K\sum_{m=1}^K \mathbf{t}_{1,l,k,m}^{\left(j \right)}\right]_n \!\!\!\!\!-\!\! \sum_{n'\ne n}^N e^{-j \theta_{l,n'}} \! \left[\sum_{k=1}^K\sum_{m=1}^K \mathbf{T}_{2,l,k,m}^{\left(j \right)}\right]_{n'\!,n} \!\!\!\!\!, \nonumber
		\\
		&\mathbf{t}_{1,l,k,m}^{\left(j \right)}= \left( \!\! \left(z_{km}^{\left(j+1 \right)} \!\!+\!\! z_{d,km}^{\left(j \right)}\right)^* \!\!\! \mathbf{h}_{\text{SIM-}k}^H \mathbf{U}_l^{\left(j \right)} \right) \! \odot \! \left(\mathbf{V}_l^{\left(j \right)} \mathbf{w}_{1,m} \rho_m^{\left(j+1 \right)}\right)^T \!\!\!, \nonumber
		\\
		&\mathbf{T}_{2,l,k,m}^{\left(j \right)}=\left(\mathbf{t}_{2,l,k,m}^{\left(j \right)} \left(\mathbf{t}_{2,l,k,m}^{\left(j \right)}\right)^H\right)^T,\nonumber
		\\
		&\mathbf{t}_{2,l,k,m}^{\left(j \right)}=\mathrm{diag}\left(\mathbf{h}_{\text{SIM-}k}^H \mathbf{U}_l^{\left(j \right)} \right) \mathbf{V}_l^{\left(j \right)} \mathbf{w}_{1,m} \rho_m^{\left(j+1 \right)},\nonumber 
		\\
		&\mathbf{U}_l^{\left(j \right)}=\boldsymbol{\Theta}_{L}^{\left(j \right)} \mathbf{W}_{L} \cdots \boldsymbol{\Theta}_{l+1}^{\left(j \right)} \mathbf{W}_{l+1},\nonumber
		\\
		&\mathbf{V}_l^{\left(j \right)}= \mathbf{W}_{l} \boldsymbol{\Theta}_{l-1}^{\left(j \right)} \mathbf{W}_{l-1} \cdots \boldsymbol{\Theta}_{2}^{\left(j \right)} \mathbf{W}_{2} \boldsymbol{\Theta}_{1}^{\left(j \right)},\nonumber
		\\
		&d_{l,n}^{\left(j \right)}=\sum_{k=1}^K\sum_{m=1}^K \left|z_{km}^{\left(j+1 \right)} \!\!+\!\! z_{d,km}^{\left(j \right)}\right|^2 -2\Re\left( \sum_{m\ne n}^N e^{j \theta_{l,m}} t_{l,m}^{\left(j \right)} \right). \nonumber
\end{align}
We can see that $d_{l,n}^{\left(j \right)}$ keep constant and $\Re\left(t_{l,n}^{\left(j \right)} e^{j \theta_{l,n}}\right) = \left|t_{l,n}^{\left(j \right)}\right| \cos\left(\arg\left( t_{l,n}^{\left(j \right)} \right)+\theta_{l,n}\right) $. Consequently, the optimal $\theta_{l,n}^{\left(j+1 \right)} $ for (P1.3.3$_{l,n}$) is given by
\begin{align} \label{eq_PhaseSolution} 
	\theta_{l,n}^{\left(j+1 \right)} = -\arg\left( t_{l,n}^{\left(j \right)} \right).
\end{align}
Based on \eqref{eq_PhaseSolution}, we can update all $LN$ phase shifts alternately and thereby can obtain  $\{\boldsymbol{\Theta}_{l}^{\left(j+1 \right)}\}$ and $\mathbf{G}^{\left(j+1 \right)}$. 

\textbf{STEP 4}: Updating $\gamma^{\left(j+1 \right)}$: \\
Given $\{z_{km}^{\left(j+1 \right)},\eta_k^{\left(j+1 \right)}\}$, $\{\rho_k^{\left(j+1 \right)}\}$, $\{\boldsymbol{\Theta}_{l}^{\left(j+1 \right)}\}$, and $\{z_{d,km}^{\left(j \right)},\eta_{d,k}^{\left(j \right)}\}$, $\gamma$ can be updated by solving
\begin{align}
	\gamma^{\left(j+1 \right)}=& \mathop{\arg\max}\limits_{\gamma}  {\cal L}_{P1.3}\left( {\{z_{km}^{\left(j+1 \right)},\eta_k^{\left(j+1 \right)},\rho_k^{\left(j+1 \right)},}\right.  \nonumber
	\\
	&\left. {\boldsymbol{\Theta}_{l}^{\left(j+1 \right)},z_{d,km}^{\left(j \right)},\eta_{d,k}^{\left(j \right)}\}}  \right),\label{eq_gammaADMM} 
\end{align}
which is equivalent to
\begin{align} \label{P3.4} 
	\mathrm{\left(P 1.3.4 \right)}:&\mathop {\max }\limits_{ \gamma } \ \gamma -\frac{\rho}{2} \sum_{k=1}^K  \left( \eta_k^{\left(j+1 \right)}-\gamma+ {\eta}_{d,k}^{\left(j \right)}  \right)^2.
\end{align}
Apparently, (P1.3.4) is a concave quadratic unconstrained maximization problem, which can be readily solved by letting the first-order derivative of the objective function w.r.t. $\gamma$ be zero. Then, $\gamma^{\left(j+1 \right)}$ is given by
\begin{align} \label{eq_gamma^i+1} 
	&\gamma^{\left(j+1 \right)}=\frac{\sum\limits_{k=1}^{K} \left(\eta_k^{\left(j+1 \right)} + {\eta}_{d,k}^{\left(i \right)} \right) \rho + 1}{\rho K}.
\end{align}

\textbf{STEP 5}: Updating $\{z_{d,km}^{\left(j+1 \right)},\eta_{d,k}^{\left(j+1 \right)}\}$: \\
$\{z_{d,km},\eta_{d,k}\}$ can be updated by
\begin{align}
	&z_{d,km}^{\left(j+1 \right)}=z_{km}^{\left(j+1 \right)}-\mathbf{h}_{\text{SIM-}k}^H \mathbf{G}^{\left(j+1 \right)} \mathbf{w}_{1,m} \rho_m^{\left(j+1 \right)}+z_{d,km}^{\left(j \right)},\label{eq_zdkm^j+1}
	\\
	&\eta_{d,k}^{\left(i+1 \right)}=\eta_{k}^{\left(j+1 \right)}- \gamma^{\left(j+1 \right)}+\eta_{d,k}^{\left(j \right)}.\label{eq_etadk^j+1}
\end{align}
	\begin{algorithm}[t]\label{Algorithm1}
	\caption{The proposed algorithm for MR maximization}
	\begin{algorithmic}[1]
		\STATE Initialize outer iteration index $i$, inner consensus-ADMM iteration index $j$,  $\rho_k^{\left(0 \right)}$, $\boldsymbol{\Theta}_{l}^{\left(0 \right)}$, $\gamma^{\left(0 \right)}$, $z_{d,km}^{\left(0 \right)}=0$, and  $\eta_{d,k}^{\left(0 \right)}=0$, for $l \in \mathcal{L}$ and $k,m \in \mathcal{K}$. 
		\REPEAT
		\STATE Compute $\{\mathbf{B}_k^{(i)}\}$, $\{c_k^{\left(i \right)}\}$, and $\gamma^{\left(i \right)}$ by $\{\rho_k^{\left(i \right)}\}$ and $\{\boldsymbol{\Theta}_{l}^{\left(i \right)}\}$. 
		\STATE Set 	$\gamma^{\left(j \right)}=\gamma^{\left(i \right)}$, $\rho_k^{\left(j \right)}=\rho_k^{\left(i \right)}$, $\boldsymbol{\Theta}_{l}^{\left(j \right)}=\boldsymbol{\Theta}_{l}^{\left(i \right)}$.
		\REPEAT		
		\STATE Update $\{z_{km}^{\left(j+1 \right)}\}$ and  $\{\eta_{k}^{\left(j+1 \right)}\}$ by Table~\ref{tab1}.	
		\STATE Update $\{\rho_k^{\left(j+1 \right)}\}$ by \eqref{eq_rho^j+1}.
		\STATE Update $\{\boldsymbol{\Theta}_{l}^{\left(j+1 \right)}\}$ by \eqref{eq_PhaseSolution}.
		\STATE Update $\gamma^{\left(j+1 \right)}$ by \eqref{eq_gamma^i+1}.
		\STATE Update $\{z_{d,km}^{\left(j+1 \right)},\eta_{d,k}^{\left(j+1 \right)}\}$ by \eqref{eq_zdkm^j+1} and \eqref{eq_etadk^j+1}.
		\STATE $j \leftarrow  j+1$.
		\UNTIL  The growth of $\min(\{\eta_{k}\})$ is less than the set threshold $\zeta_{inn}>0$, i.e., $(\min(\{\eta_{k}^{\left(j \right)}\})-\min(\{\eta_{k}^{\left(j-1 \right)}\}))/\min(\{\eta_{k}^{\left(j \right)}\}) \le \zeta_{inn}$ or the maximum number of iterations $\tau_{inn}>0$ has been reached. 
		\STATE Output $\rho_k^{\left(i+1 \right)}=\rho_k^{\left(j \right)}$ and $\boldsymbol{\Theta}_{l}^{\left(i+1 \right)}=\boldsymbol{\Theta}_{l}^{\left(j \right)}$	
		\STATE   $i \leftarrow  i+1$.
		\UNTIL  The growth of $\gamma$ is less than the set threshold $\zeta_{out}>0$, i.e., $(\gamma^{\left(i \right)}-\gamma^{\left(i-1 \right)})/\gamma^{\left(i \right)} \le \zeta_{out}$ or the maximum number of iterations $\tau_{out}>0$ has been reached. 
	\end{algorithmic}
\end{algorithm}
\subsection{Convergence and Computational Complexity Analysis}
To elaborate, the proposed algorithm is summarized in Algorithm 1. The convergence behavior of the inner consensus-ADMM under similar problem structures has been consistently observed and verified in \cite{Huang2016,Boyd2011,Hong2012,Fang2024}. Moreover, based on \eqref{eq_r_k_hat}, the outer iteration always satisfies\cite{Yu2020}
\begin{align} \label{eq_convergence} 
&r_k(\{\rho_k^{\left(i+1 \right)},\boldsymbol{\Theta}_{l}^{\left(i+1 \right)}\}) \ge 
\hat{r}_{k,i}(\{\rho_k^{\left(i+1 \right)}\!,\boldsymbol{\Theta}_{l}^{\left(i+1 \right)}\}|\{\rho_k^{\left(i \right)}\!,\! \boldsymbol{\Theta}_{l}^{\left(i \right)}\}) \! \nonumber
\\
&\ge \!  \hat{r}_{k,i}(\{\rho_k^{\left(i \right)}\!,\! \boldsymbol{\Theta}_{l}^{\left(i \right)}\}|\{\rho_k^{\left(i \right)}\!,\!\boldsymbol{\Theta}_{l}^{\left(i \right)}\}) 
\!\!
=\!\!  r_k(\{\rho_k^{\left(i+1 \right)},\boldsymbol{\Theta}_{l}^{\left(i+1 \right)}\}).
\end{align} 
Therefore, the algorithm typically converges to at least a locally optimal solution of (P1).

In general, we usually simplify the expression of computational complexity by omitting the lower order terms and constant coefficients\cite{Boyd2004}. Note that Algorithm 1 is computationally efficient, since all optimized variables are updated by closed-form expressions or the simple bisection search. By \eqref{eq_r_k_hat}, \eqref{eq_b11}, \eqref{eq_b12}, and \eqref{eq_b22}, the computational complexity of step 3 in Algorithm 1 is mainly dominated by computing $\mathbf{h}_{\text{SIM-}k}^H \mathbf{G} \mathbf{w}_{1,m} \rho_m$, for $k,m \in \mathcal{K}$,  of which the computational complexity is $\mathcal{O}(K^2 N^2)$. Similarly, the computational complexity of step 6 is $\mathcal{O}((K^2 N^2)\log_2(1/\varepsilon ))$, where $\varepsilon$ stands for the accuracy of bisection searching. Since $\mathbf{h}_{\text{SIM-}k}^H \mathbf{G}^{(j)} \mathbf{w}_{1,m}$, for $k,m \in \mathcal{K}$ has been computed in step 6, the computational complexity of step 7 is $\mathcal{O}(K^2\log_2(1/\varepsilon ))$. Likewise, the computational complexity of step 8, 9, and 10 are $\mathcal{O}(L N^3 K^2)$, $\mathcal{O}(K)$, and $\mathcal{O}(K^2 N^2)$, respectively. Finally, the overall computational complexity of Algorithm 1 can be expressed as $\mathcal{O}(I_{out}( I_{inn}( (K^2 N^2)\log_2(1/\varepsilon ) + K^2\log_2(1/\varepsilon )+ L N^3 K^2 + K^2 N^2 )+ K^2 N^2))$, where $I_{inn}$ and $I_{out}$ denote the numbers of iterations required for inner consensus-ADMM iteration and outer iteration, respectively.

\section{Solving Algorithm for GMR Maximization}
\subsection{GMR Maximization Algorithm Design}
Again, assuming $\{\rho_k^{\left(i \right)}\}$ and $\{\boldsymbol{\Theta}_{l}^{\left(i \right)}\}$ are feasible solutions outputted in the $i$-th round of iteration, we can then obtain the corresponding $\{r_k^{\left(i \right)}\}$.  As shown in \cite{Yu2022,Chen2024,Yu2024,Tuan2022,Zhu2023a}, for the given   $\{\rho_k^{\left(i \right)}\}$ and $\{\boldsymbol{\Theta}_{l}^{\left(i \right)}\}$, the linearized function of \eqref{P2} can be expressed as
\begin{align} \label{PGM_liner}
	 \xi^{\left(i \right)} \left(\{\rho_k,\boldsymbol{\Theta}_{l}\} \right) = \sum_{k=1}^K  \nabla f_{GM,k}^{\left(i \right)} r_k + \frac{K-1}{K} f_{GM}\left(\{r_k^{\left(i \right)}\}\right),
\end{align}
in which $f_{GM}\left(\{r_k^{\left(i \right)}\}\right) \triangleq \left( \prod_{k=1}^K r_k^{\left(i \right)} \right)^{1/K}$ and $\nabla f_{GM,k}^{\left(i \right)}=f_{GM}\left(\{r_k^{\left(i \right)}\}\right)/ K r_k^{\left(i \right)} \ge 0$ represents the gradient of  $f_{GM}\left(\{r_k\}\right)$  w.r.t $\{r_k\}$ at feasible solution $\{r_k^{\left(i \right)}\}$. Therefore, at the $i$-th round, we can seek  $\{\rho_k^{\left(i+1 \right)}\}$ and $\{\boldsymbol{\Theta}_{l}^{\left(i+1 \right)}\}$ by solving
\begin{align} \label{P2.1}
	\mathrm{\left(P 2.1\right)}: & \mathop{\max}\limits_{ \{\rho_k,\boldsymbol{\Theta}_{l}\}} \mathop     {\xi^{\left(i \right)} \left(\{\rho_k,\boldsymbol{\Theta}_{l}\} \right)} 
	\\ \mbox{s.t.}\quad
	&\eqref{P1b}, \eqref{P1c}, \eqref{P1d}.   \tag{\ref{P2.1}{a}}
\end{align}
 Unfortunately, (P2.1) can still not be directly addressed due to the complex coupling. With the assistance of \eqref{eq_r_k_hat}, (P2.1) can be transformed into two sub-problems relative to $\{\rho_k\}$ and $\{\boldsymbol{\Theta}_{l}\}$, respectively, and then we can update both sub-problems alternately. Next, we will show the details of the alternative iteration procedure.

\textbf{STEP 1}: Updating $\{\rho_k^{\left(i+1 \right)}\}$: \\
Given $\{\boldsymbol{\Theta}_{l}^{\left(i \right)}\}$, $\{\rho_k\}$ can be updated by solving
\begin{align} \label{P5.1}
	&\mathrm{\left(P 2.1.1\right)}\!\!:\!\! \{\rho_k^{\left(i+1\right)}\}= \mathop{\arg\max}\limits_{ \{\rho_k\}} \mathop     { \sum_{k=1}^{K}  \frac{\nabla f_{GM,k}^{\left(i \right)}}{\ln2}
		\left(	c_k^{\left(i \right)}-2 \Re \left( b_{k,12}^{\left(i \right)} \right.   \right. } \nonumber
	\\ &\left. \left. \mathbf{h}_{\text{SIM-}k}^H \mathbf{G}^{\left(i \right)} \mathbf{w}_k^{1}  \rho_k \right)-\! b_{k,22}^{\left(i \right)} \!\! \sum\limits_{m=1}^{K} \! \left|  \mathbf{h}_{\text{SIM-}k}^H \mathbf{G}^{\left(i \right)} \mathbf{w}_m^{1} \rho_m \right|^2 \! \right)  \\&\mbox{s.t.} \quad
	\eqref{P1b}, \eqref{P1c}.  \tag{\ref{P5.1}{a}}
\end{align}
Ignoring the non-negative constraint \eqref{P1c}, (P2.1.1) also becomes a convex QCQP-1 that can be solved in closed-form by multiplier method. 
Then, $\rho_k^{\left(i+1 \right)}$ can be given by
\begin{equation}\label{eq_rho^i+1}
	\begin{aligned}
		\rho_k^{\left(i+1 \right)} =
		\begin{cases}
			f_{k,\mu}\left(0 \right), \ \text{if} \; \sum_{k=1}^K \left(  f_{k,\mu}\left(0 \right) \right)^2 \le P_{max},
			\\ 
			f_{k,\mu}\left( \hat{\mu} \right), \ \text{otherwise}.
		\end{cases} 
	\end{aligned}
\end{equation}
where  
\begin{align} 
	f_{k,\mu}\left( \mu \right) \!\!=\!\!\frac{ -\nabla f_{GM,k}^{\left(i \right)} \Re\left(b_{k,12}^{\left(i \right)} \mathbf{h}_{\text{SIM-}k}^H \mathbf{G}^{\left(i \right)} \mathbf{w}_{1,k}\right) } {\sum\limits_{m=1}^{K} \nabla f_{GM,m}^{\left(i \right)} b_{m,22}^{\left(i \right)} \left| \mathbf{h}_{\text{SIM-}m}^H \mathbf{G}^{\left(i \right)} \mathbf{w}_{1,k}\right|^2  \!+\! \ln2 \mu}, \nonumber
\end{align}
for $ \mu \ge  0$, where $ \hat{\mu} >  0$ is the optimal dual variable when $\sum_{k=1}^K \left(  f_{k,\mu}\left(0 \right) \right)^2 > P_{\max}$. By substituting \eqref{eq_b12} and \eqref{eq_b22} into  \eqref{eq_rho^i+1}, we can see that  $\{\rho_k^{\left(i+1 \right)}\}$ is always non-negative. This also implies that constraint \eqref{P1c} is always satisfied.

\textbf{STEP 2}: Updating $\{\boldsymbol{\Theta}_{l}^{\left(i+1 \right)}\}$: \\
Given $\{\rho_k^{\left(i+1 \right)}\}$, $\{\boldsymbol{\Theta}_{l}\}$ can be updated by solving
\begin{align} \label{P5.2}
	&\mathrm{\left(P 2.1.2\right)}\!\!:\!\! \{\boldsymbol{\Theta}_{l}^{\left(i+1 \right)}\}= \mathop{\arg\max}\limits_{ \{\boldsymbol{\Theta}_{l}\}} \mathop     { \sum_{k=1}^{K} \frac{\nabla f_{GM,k}^{\left(i \right)}}{\ln2}
		\left(	c_k^{\left(i \right)}-2 \Re \left( b_{k,12}^{\left(i \right)} \right.   \right. } \nonumber
	\\ &\left. \left. \mathbf{h}_{\text{SIM-}k}^H \mathbf{G} \mathbf{w}_k^{1}  \rho_k^{\left(i+1 \right)} \!\! \right) \!\!-\! b_{k,22}^{\left(i \right)} \!\! \sum\limits_{m=1}^{K} \! \left|  \mathbf{h}_{\text{SIM-}k}^H \mathbf{G} \mathbf{w}_m^{1} \rho_m^{\left(i+1 \right)} \right|^2 \! \right) 
	\\
	&\mbox{s.t.} \quad 0 \le \theta_{l,n} \le 2\pi. \tag{\ref{P5.2}{a}}
\end{align}
Similar to (P1.3.3), the phase shift of the $n$-th meta-atom on the $l$-th metasurface layer $\theta_{l,n}$ can be updated by solving 
\begin{align} \label{P5.2ln} 
	\mathrm{\left({P 2.1.2}_\emph{l,n}\right)}: &\mathop{\max}\limits_{ \theta_{l,n} } \mathop 
	-2\Re\left(q_{l,n}^{\left(i \right)} e^{j \theta_{l,n}}\right)
	\\
	&\mbox{s.t.} \quad 0 \le \theta_{l,n} \le 2\pi. \tag{\ref{P5.2ln}{a}} \label{P5.2lna}
\end{align}
where
\begin{align}
	&q_{l,n}^{\left(i \right)}=\left[\sum_{k=1}^K \mathbf{q}_{1,l,k}^{\left(i \right)} \right]_n \!\!-\!\! \sum_{n'\ne n}^N e^{-j \theta_{l,n'}} \! \left[\sum_{k=1}^K\sum_{m=1}^K \mathbf{Q}_{2,l,k,m}^{\left(j \right)}\right]_{n',n} \!\!\!, \nonumber
	\\
	&\mathbf{q}_{1,l,k}^{\left(i \right)}= 	\left( \!\!\nabla f_{GM,k}^{\left(i \right)} b_{k,12}^{\left(i \right)} \mathbf{h}_{\text{SIM-}k}^H \mathbf{U}_l^{\left(i \right)}\right) \! \odot \! \left(\mathbf{V}_l^{\left(j \right)} \mathbf{w}_{1,k} \rho_k^{\left(i+1 \right)}\right)^T, \nonumber
	\\
	&\mathbf{Q}_{2,l,k,m}^{\left(i \right)}=\nabla f_{GM,k}^{\left(i \right)} b_{k,22}^{\left(i \right)}\left(\mathbf{q}_{2,l,k,m}^{\left(i \right)} \left(\mathbf{q}_{2,l,k,m}^{\left(i \right)}\right)^H\right)^T, \nonumber
	\\ 
	&\mathbf{q}_{2,l,k,m}^{\left(i \right)}=\mathrm{diag}\left(\mathbf{h}_{\text{SIM-}k}^H \mathbf{U}_l^{\left(i \right)} \right) \mathbf{V}_l^{\left(i \right)} \mathbf{w}_{1,m} \rho_m^{\left(i+1 \right)}.\nonumber 
\end{align}
Consequently, the optimal $\theta_{l,n}^{\left(i+1 \right)} $ for (P2.1.2$_{l,n}$) is given by
\begin{align} \label{eq_PhaseSolutionGM} 
	\theta_{l,n}^{\left(i+1 \right)} = \pi -\arg\left( q_{l,n}^{\left(i \right)} \right).
\end{align}
\begin{algorithm}[!t]\label{Algorithm2}
	\caption{The proposed algorithm for GMR maximization}
	\begin{algorithmic}[1]
		\STATE Initialize iteration index $i=0$, $\rho_k^{\left(0 \right)}$, and $\boldsymbol{\Theta}_{l}^{\left(0 \right)}$
		\REPEAT	
		\STATE Compute $\{\mathbf{B}_k^{(i)}\}$ and $\{c_k^{\left(i \right)}\}$ by $\{\rho_k^{\left(i \right)}\}$ and $\{\boldsymbol{\Theta}_{l}^{\left(i \right)}\}$. 
		\STATE Given $\{\boldsymbol{\Theta}_{l}^{\left(i \right)}\}$, update $\{\rho_k^{\left(i+1 \right)}\}$ by \eqref{eq_rho^i+1}.
		\STATE Given $\{\rho_k^{\left(i+1 \right)}\}$, update $\{\boldsymbol{\Theta}_{l}^{\left(i +1\right)}\}$ by \eqref{eq_PhaseSolutionGM}.
		\STATE  $i \leftarrow  i+1$.
		\UNTIL  The growth of $f_{GM}\left(\{r_k\}\right)$ is less than the set threshold $\zeta_{GM}>0$ or the maximum number of iterations has been reached. 
	\end{algorithmic}
\end{algorithm}

Algorithm 2 provides the pseudo-code for the proposed steepest descent procedure to solve (P2.1). The iterations of \eqref{eq_rho^i+1} and \eqref{eq_PhaseSolutionGM} obtain a descent direction by finding a better feasible point for the non-convex problem (P2). This surrogate-based strategy is similar to the Frank-and-Wolfe method\cite{Yu2022}, which guarantees convergence to a stationary point under mild conditions. According to \cite{Yu2022,Zhu2023a,Nasir2022}, by repeatedly alternating the iterations of \eqref{eq_rho^i+1} and \eqref{eq_PhaseSolutionGM}, the algorithm typically converges to at least a locally optimal solution of (P2). The overall computational complexity of Algorithm 2 can be expressed as $\mathcal{O}(I_{GM}((K^2 N^2)(\log_2(1/\varepsilon)+1) + L N^3 K^2 ))$, where $I_{GM}$ denotes the numbers of iterations required for convergence. 
\subsection{Solving Algorithm for SR Maximization}
By (P2.1), we can observe that both $\nabla f_{GM,k}^{(i)}$ and $ f_{GM}(\{r_k^{(i)}\})$ are constant terms in the $i$-th iteration, because both of which are determined by $\{r_k^{\left(i \right)}\}$, not $\{r_k\}$.  Therefore, if we let $\nabla f_{GM,k}^{\left(i \right)}=1$ and $f_{GM}(\{r_k^{\left(i \right)}\})=0$ in each iteration,  the optimization objective in (P2.1)  becomes $\sum_{k=1}^K r_k$, representing the system SR. Then, Algorithm 2 can be directly applied to the SR maximization. 
\begin{Remark}
For the three problems of MR maximization, GMR maximization, and SR maximization, we have derived the corresponding closed-form solutions. Notably, the algorithmic complexity decreases progressively from MR to GMR and then to SR. MR maximization is appropriate for scenarios where user fairness is a primary consideration, as it aims to improve the performance of the user with the lowest rate. GMR maximization offers a balanced trade-off between fairness and throughput, making it suitable for environments with moderate user heterogeneity. SR maximization focuses on maximizing the overall system throughput and is generally more effective when all users experience the same degree of large-scale fading.
\end{Remark}
\section{Numerical Evaluation}
\subsection{Evaluation Scenarios and Parameters}
We consider a 3-dimension (x-y-z coordinate system) simulation environment and the unit of distance is 1 meter. Specifically, the BS is equipped with a uniform linear array (ULA) with few antennas and each layer of the SIM arrangement is identically equipped with a uniform planar array (UPA) containing $N=N_xN_z$ elements, where $N_x$ and $N_z$ are the numbers of elements along x-axis and z-axis, respectively. Next we assume that $N_x=N_z$. Importantly, the BS transmit antennas along z-axis are integrated with the SIM to facilitate beamforming design in the EM wave domain~\cite{An2023a}.  The SIM is deployed on the x-z plane with the reference meta-atom located at $(0,0,0)$, and all $K$ users are randomly distributed in the user cluster centered at $(0,60,0)$ with radius $50$m. 

For the channel from the SIM to the $k$-th user, path loss coefficient is equal to   $\beta_{\text{SIM-}k}=10^{({G_\mathrm{BS}} + {G_{k}} - 33.05)/10}d_{\text{SIM-}k}^{-\alpha_{\text{SIM-}k}}$, where  $G_{\mathrm{BS}}$ and $G_{k}$ represent the antenna gains of BS and user $k$, respectively, $d_{\text{SIM-}k}$ represents the link distance from the SIM to the $k$-th user, and $\alpha_{\text{SIM-}k}$ represents the path loss exponent. Denote the spatial correlation matrix of the SIM w.r.t. the $k$-th user by $\mathbf{R}_{\text{SIM-}k} \in \mathbb{C}^{N \times N}$ and there is $[\mathbf{R}_{\text{SIM-}k}]_{n,n'}=e^{j \pi (n-n') \sin \psi_{k} \sin \varphi_{k} }$ in which $\psi_{k}$ and $\varphi_{k}$ are the azimuth and elevation angle w.r.t. user $k$, respectively.   
Assume  the channel from the SIM to each user experience Rayleigh fading,  then we have $\mathbf{h}_{\text{SIM-}k}^H\sim \mathcal{CN}(0,\beta_{\text{SIM-}k}\mathbf{R}_{\text{SIM-}k})$. Unless otherwise stated later, Table~\ref{parameters} demonstrates all the parameter settings. In the following legend, {SIM-$N \times L$} means the SIM has $L$ layers of metasurfaces and there are $N$ meta-atoms on each layer. 
\begin{table}[!t]\footnotesize
	\centering
	\caption{Simulation Parameters}
	\begin{tabular}{cccc} \midrule[1pt]
		Parameter & Value\\ \midrule[0.5pt]
		$K$, $M$ & 4, 4 \\
		$P_{\max}$ & 20 dBm \\
		$\sigma _{k}^2$ & $-96$ dBm \\
		$\lambda_c $ & 0.0107m (28 GHz)\cite{An2023a}\\
        ${\alpha_{\text{SIM-}k}} $  &  3\\
		 ${G_\mathrm{BS}}$, ${G_{k}}$, $\forall k$ & 5 dBi, 0 dBi \cite{Nadeem2020} \\
		$\rho$ & 100 \\
		$\zeta_{inn}$, $\zeta_{out}$, $\zeta_{GM}$ & $10^{-4}$, $10^{-5}$, $10^{-5}$ \\
		$\tau_{inn}$, $\tau_{out}$, $\tau_{GM}$ & $5000$, $8000$, $8000$\\
		$d_l$, $d_w$, $d_t$ & $0.5\lambda_c $, $0.5\lambda_c $, $5\lambda_c $ \cite{An2024,An2025}
		\\\midrule[1pt]
	\end{tabular}
	\label{parameters}
\end{table}

\subsection{Convergence of the Proposed Algorithms}
\begin{figure}[!t]
	\centering
	\subfigure[Convergence behavior of Algorithm 1.]{
		\includegraphics[width=0.44\textwidth]{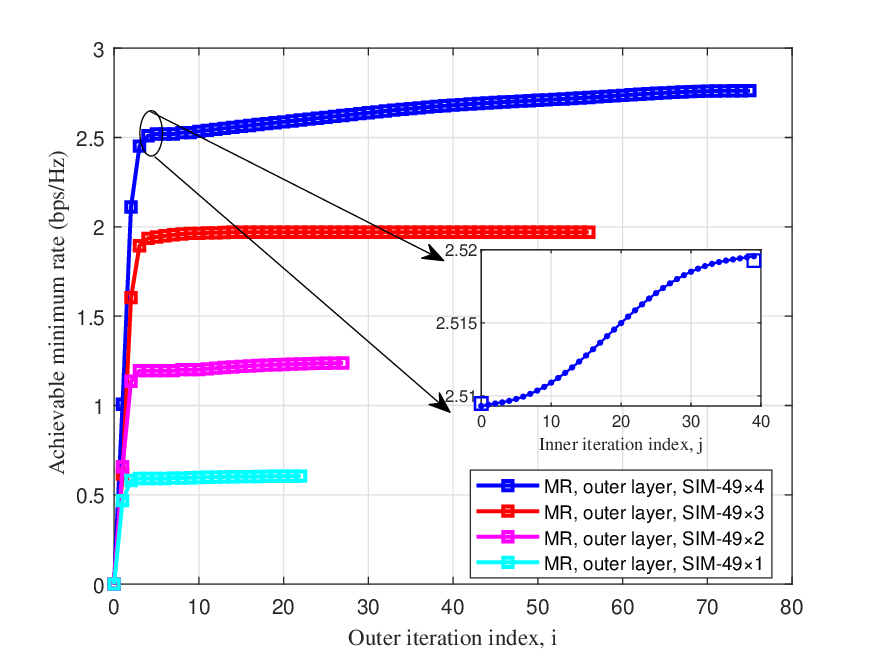}
		\label{ConvergenceAlg1Outer}
	}%
	\\
	\subfigure[Convergence behavior of Algorithm 2.]{
		\includegraphics[width=0.44\textwidth]{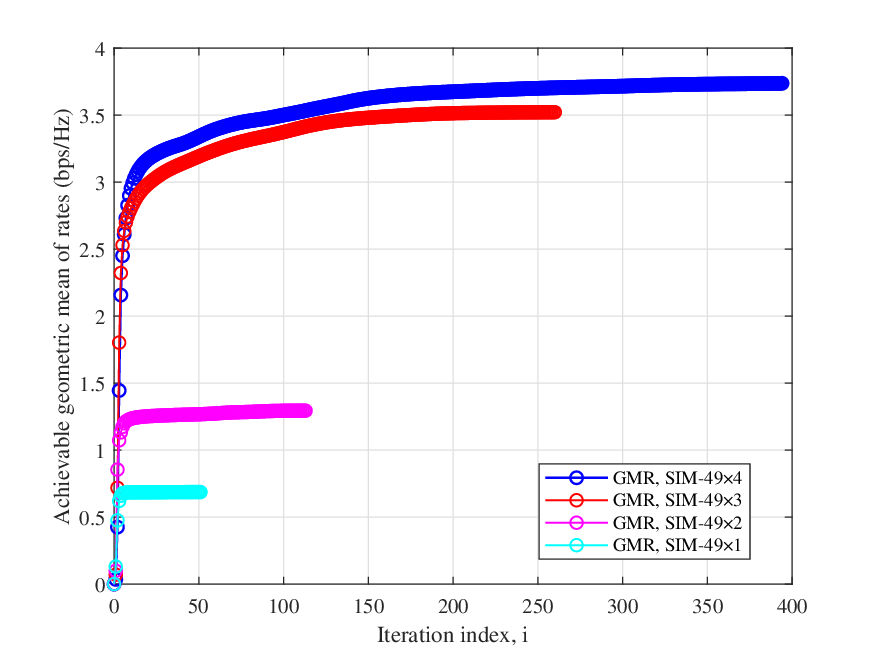}
		\label{ConvergenceAlg2}
	}%
	\centering
	\caption{Convergence behavior with  $M=K=6$ and $P_{\max}=10$dBm.}
	\label{Convergence_behaviour}
\end{figure}

Fig.~\ref{Convergence_behaviour} illustrates the convergence behavior of the proposed optimization algorithms. Both Algorithm 1 and Algorithm 2 consistently converge to stable solutions under various parameter settings, demonstrating the robustness and effectiveness of the proposed framework. Specifically, Fig.~\ref{ConvergenceAlg1Outer} presents the overall convergence process of Algorithm 1. In this figure, the horizontal axis corresponds to the outer iteration index, while the vertical axis indicates the minimum user rate. The results clearly confirm the stability and monotonic improvement of the outer-loop iterations, i.e., \eqref{eq_convergence}. To provide further insight, we zoom in on the MR performance of the SIM-49$\times$4 configuration between the 4th and 5th outer iterations. As shown, the convergence trajectory of the inner consensus-ADMM subproblem can be clearly observed, verifying the efficiency of the inner-layer updates in achieving accurate solutions within a few iterations. Fig.~\ref{ConvergenceAlg2} illustrates the convergence process of Algorithm 2, where a similar pattern of rapid and stable convergence is observed. Unlike Algorithm 1, Algorithm 2 adopts a single-layer iterative approach, resulting in a smoother convergence trajectory. In addition, as the number of SIM layers increases, the feasible space of the optimization variables expands, requiring more iterations to achieve convergence. Nevertheless, increasing the number of SIM layers also leads to improved MR and GMR performance, due to the higher DoF available for approximating the desirable beamforming matrix.
\subsection{Rate Fairness Performance Comparisons}
\begin{figure}[!t]
	\centering
	\subfigure{\includegraphics[width=0.44\textwidth]{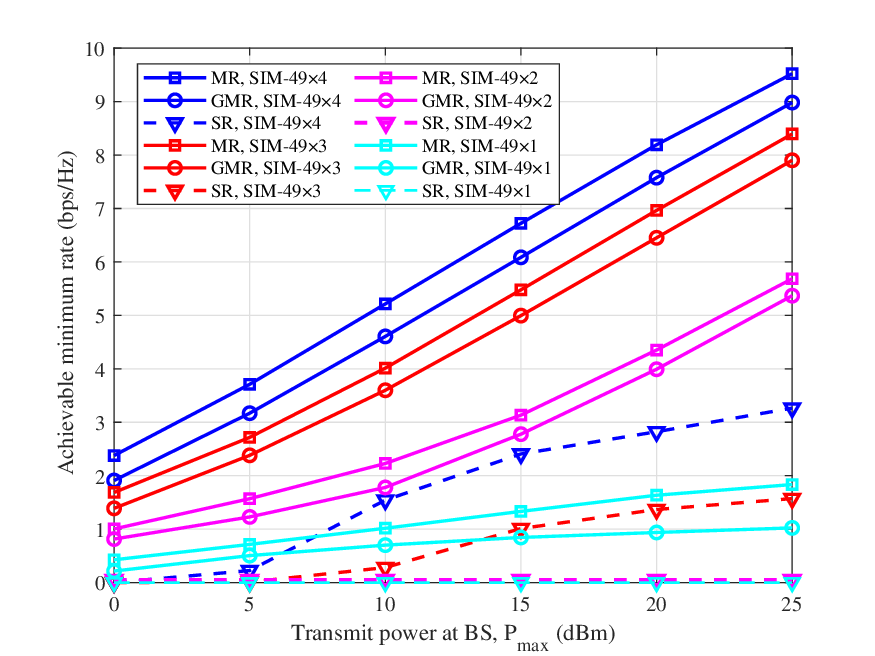}}
	\caption{Achievable MR versus the maximum allowable transmit power $P_{\max}$.}
	\label{PlotMaxminPower}
\end{figure}
\begin{figure}[!t]
	\centering
	\subfigure{\includegraphics[width=0.44\textwidth]{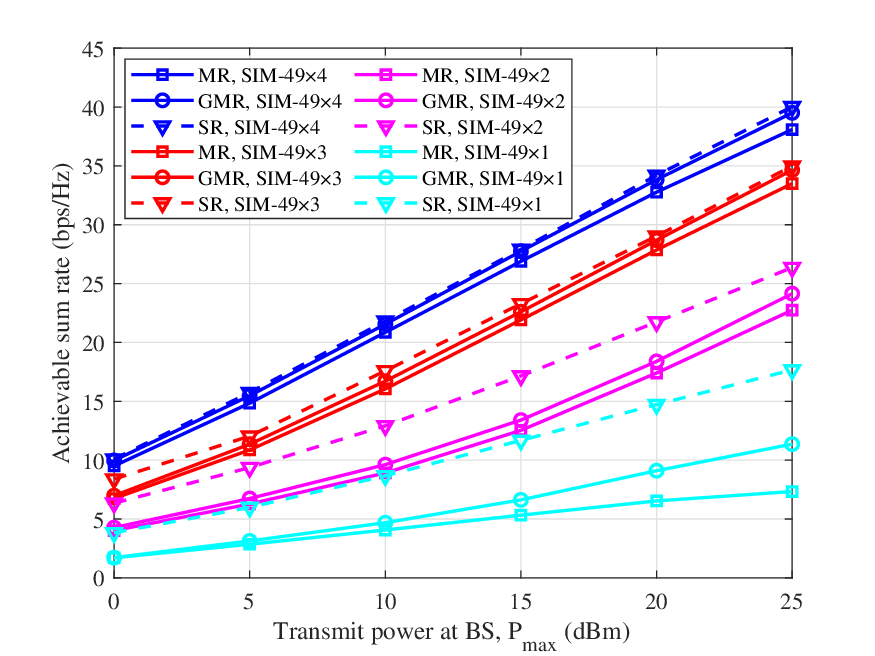}}
	\caption{Achievable SR versus the maximum allowable transmit power $P_{\max}$.}
	\label{PlotSumPower}
\end{figure}

Fig.~\ref{PlotMaxminPower} depicts the achievable MR versus the maximum allowable transmit power $P_{\max}$. With the same parameter setting, MR maximization always achieves the highest MR, whereas SR maximization yields the worst performance, even zero rate when $L=1,2$. Also, we can see that increasing the number of SIM layers enhances the MR. However, it is important to note that the increase in MR when the number of SIM layers $L$ rises from 1 to 2 is more significant than the gain observed when increasing $L$ from 2 to 3. It indicates that increasing $L$ can enhance the MR, but diminishing return exists. Fig.~\ref{PlotSumPower} presents achievable SR versus $P_{\max}$. In contrast to Fig.~\ref{PlotMaxminPower}, SR maximization achieves the best performance, while MR maximization performs the worst. Therefore, in both scenarios, we see that GMR maximization offers a more balanced rate performance than MR maximization and SR maximization, effectively representing the trade-off of rate fairness and overall rate  performance.  

\begin{table}[!t]
	\scriptsize
	\setlength\tabcolsep{3pt}
	\renewcommand\arraystretch{1.1}
	\caption{Rate Standard Deviations of Users}\label{tab:PlotVariancePower}
	\centering
	{
		\begin{tabular}{|c|c|c|c|c|c|}
			\cline{1-6} 
			\diagbox{Schemes}{$P_{\max}$} & 0dBm & 5dBm & 10dBm & 15dBm & 20dBm 
			\\
			\hline\hline
			MR, SIM-49$\times$4 & 0 &  0  & 0  &  0 & 0    
			\\
			\hline
			GMR, SIM-49$\times$4 &  0.48 &  0.58  & 0.64  &  0.69 &    0.72
			\\
			\hline
			SR, SIM-49$\times$4 & 2.43  &  2.81  & 3.06  &  3.29 &   3.67
			\\
			\hline\hline
			MR, SIM-49$\times$3 & 0  &   0 &  0 & 0  &   0
			\\
			\hline
			GMR, SIM-49$\times$3 & 0.38  &  0.50  &  0.58 & 0.67  &  0.72 
			\\
			\hline
			SR, SIM-49$\times$3 & 2.33  &  2.62  & 2.95  &  3.46 &   4.07
			\\
			\hline\hline
			MR, SIM-49$\times$2 &  0 &  0  &  0 & 0  &   0
			\\
			\hline
			GMR, SIM-49$\times$2 & 0.37  &  0.55  &  0.67 & 0.76  &  0.78
			\\
			\hline
			SR, SIM-49$\times$2 & 2.02  &  2.64  & 2.89  &  3.46 &   4.20
			\\
			\hline\hline
			MR, SIM-49$\times$1 & 0  &  0  & 0  & 0  &   0
			\\
			\hline
			GMR, SIM-49$\times$1 & 0.28  &  0.35  &  0.54 & 0.89  &   1.48
			\\
			\hline
			SR, SIM-49$\times$1 &  1.71 &  2.04  & 2.63  &  3.47 &  4.29
			\\
			\hline
	\end{tabular}}
\end{table}

\begin{table}[!t]
	\scriptsize
	\setlength\tabcolsep{3pt}
	\renewcommand\arraystretch{1.1}
	\caption{The Ratio of the Minimum Rate to the Maximum  Rate}\label{tab:PlotMinRMaxPower}
	\centering
	{
		\begin{tabular}{|c|c|c|c|c|c|}
			\cline{1-6} 
			\diagbox{Schemes}{$P_{\max}$} & 0dBm & 5dBm & 10dBm & 15dBm & 20dBm 
			\\
			\hline\hline
			MR, SIM-49$\times$4 &  0.9998 &  0.9999  & 0.9999 &  0.9999 &    0.9999  
			\\
			\hline
			GMR, SIM-49$\times$4 & 0.6445  &  0.7099  & 0.7623  &  0.7981 &   0.8215
			\\
			\hline
			SR, SIM-49$\times$4 & 0  &  0.0210 & 0.1755  &  0.2928 &   0.3653
			\\
			\hline\hline
			MR, SIM-49$\times$3 & 0.9999  & 0.9999  &  0.9999  & 0.9999 & 0.9999
			\\
			\hline
			GMR, SIM-49$\times$3 &  0.6238 &  0.6851 &  0.7358  & 0.7718 & 0.7978
			\\
			\hline
			SR, SIM-49$\times$3 & 0  & 0.0022  &  0.0590  & 0.1358 & 0.1755
			\\
			\hline\hline
			MR, SIM-49$\times$2 & 0.9998  & 0.9999  & 0.9999   & 0.9999 & 0.9999
			\\
			\hline
			GMR, SIM-49$\times$2 & 0.5236  & 0.5153  &  0.5500  & 0.6262 & 0.6998
			\\
			\hline
			SR, SIM-49$\times$2 & 0  &  0 &  0  & 0 & 0
			\\
			\hline\hline
			MR, SIM-49$\times$1 &  0.9997 & 0.9998  &  0.9998  & 0.9998 & 0.9998
			\\
			\hline
			GMR, SIM-49$\times$1 & 0.2693  & 0.2567  &  0.3306  & 0.3902 & 0.4017
			\\
			\hline
			SR, SIM-49$\times$1 & 0  &  0 &   0 & 0 & 0
			\\
			\hline
	\end{tabular}}
\end{table}

In Table~\ref{tab:PlotVariancePower}, we present the rate standard deviations of all users through three rate optimization algorithms. Note that, to avoid statistical fluctuation, referring to the range of user rates, we herein approximately deem the rate standard deviation as zero if it is less than $10^{-3}$ bps/Hz. With the same $P_{\max}$, MR maximization nearly has zero deviation, followed by GMR maximization, and SR maximization achieves the largest standard deviation. It indicates that MR maximization is the fairest solution and SR maximization has the least fairness.      
As $P_{\max}$ increases, the rate standard deviations of MR  maximization remains zero, indicating the minimum rate variation. GMR maximization shows a slight increase in standard deviation but stays at a very  low level. In contrast, SR maximization exhibits greater fluctuations, with standard deviations ranging from 1.71 to 4.29, reflecting a less stable rate distribution.

Table~\ref{tab:PlotMinRMaxPower} illustrates the ratio of the minimum rate to the maximum rate versus $P_{\max}$. This ratio, ranging from 0 to 1, quantifies rate fairness, with 1 indicating perfect fairness and lower values reflecting increased disparity. Simulation results demonstrate that MR maximization and GMR maximization achieve far higher ratios than SR maximization, underscoring their superiority in fairness enhancement. As $P_{\max}$ increases, the ratio for MR maximization remains near 1 across different SIM layer configurations, demonstrating its robustness in maintaining fairness.  However, some ratios for SR maximization is zero, which indicates the inability to attain a satisfactory rate.
\subsection{Performance Comparison with Existing Works}
\begin{figure}[!t]
	\centering
	\subfigure{\includegraphics[width=0.44\textwidth]{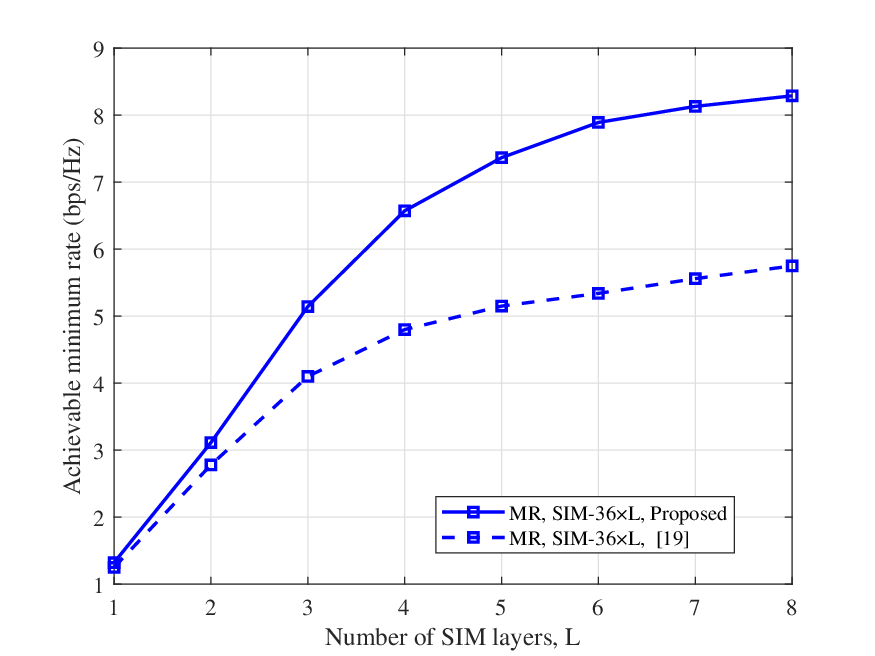}}
	\caption{MR comparison of the proposed Algorithm 1 and  \cite{Ginige2025} with same parameters setups as in \cite{Ginige2025}.}
	\label{PlotComparison1}
\end{figure} 
\begin{figure}[!t]
	\centering
	\subfigure{\includegraphics[width=0.44\textwidth]{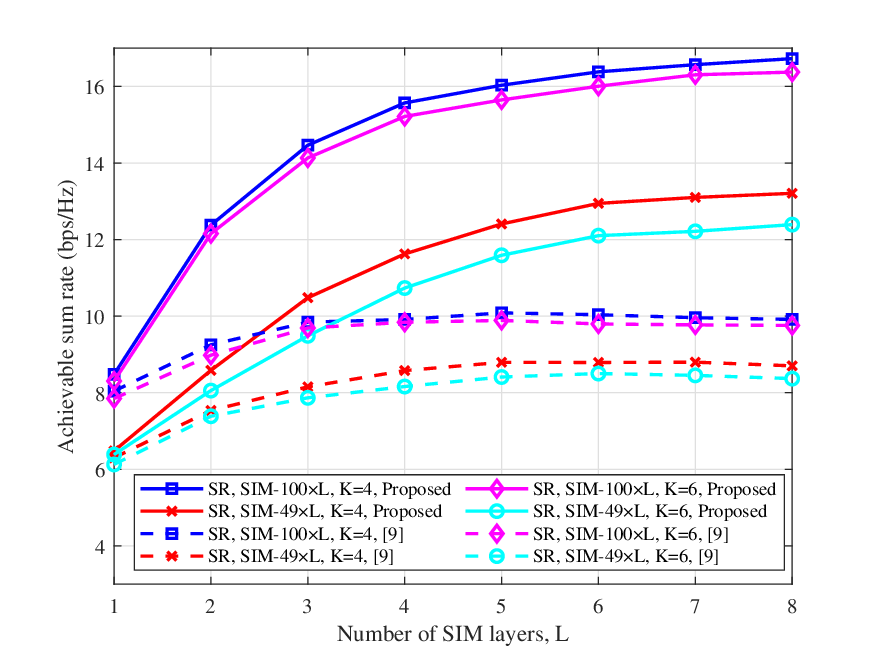}}
	\caption{SR comparison of the proposed Algorithm 2 and \cite{An2023} with same parameters setups as in \cite{An2023}.}
	\label{PlotComparison}
\end{figure} 
To further evaluate the effectiveness of the proposed Algorithm 1, we compare its MR performance with \cite{Ginige2025} in Fig.~\ref{PlotComparison1}\footnote{The code of the proposed algorithms for MR and SR maximization are available at: https://github.com/JunjieFangXJTU/Proposed}. For a fair comparison, we adopt the same parameters setups as in \cite{Ginige2025}.  The superior performance of Algorithm 1 primarily arises from two factors. First, while \cite{Ginige2025} typically focus on MSINR maximization, our approach directly optimizes the MR metric, which better captures user fairness at the rate level. Second, we derive a closed-form solution for the SIM phase shifts, eliminating the dependence on iterative gradient descent methods and thereby enhancing solution accuracy. It can be observed that as $L$ increases, the performance gap between the proposed algorithm and \cite{Ginige2025} becomes increasingly significant, which further validates the second reason discussed above. Similarly, to assess the performance of the proposed Algorithm 2 for SR maximization, Fig.~\ref{PlotComparison} presents a comparison between Algorithm 2 and \cite{An2023}. Since most existing studies on SR maximization generally employ water-filling or gradient descent-based methods, we adopt the widely recognized classical work \cite{An2023} as the benchmark\footnote{The code of \cite{An2023} is sourced from: https://github.com/JianchengAn/SIM-2-MUMISO}. 
For a fair comparison, we here also set the identical configuration parameters as in \cite{An2023}.  As $L$ increases, the proposed algorithm demonstrates gradually increasing SR gains, e.g.,    outperforming the benchmark algorithm by {35.36\%} at {SIM-100$\times$2}, {66.65}\% at {SIM-100$\times$6}, and {67.83}\% at {SIM-100$\times$8} for $K=6$. 

\subsection{Rate Comparison With Digital Beamforming}
\begin{figure}[!t]
	\centering
	\subfigure[MR maximization.]{
		\includegraphics[width=0.44\textwidth]{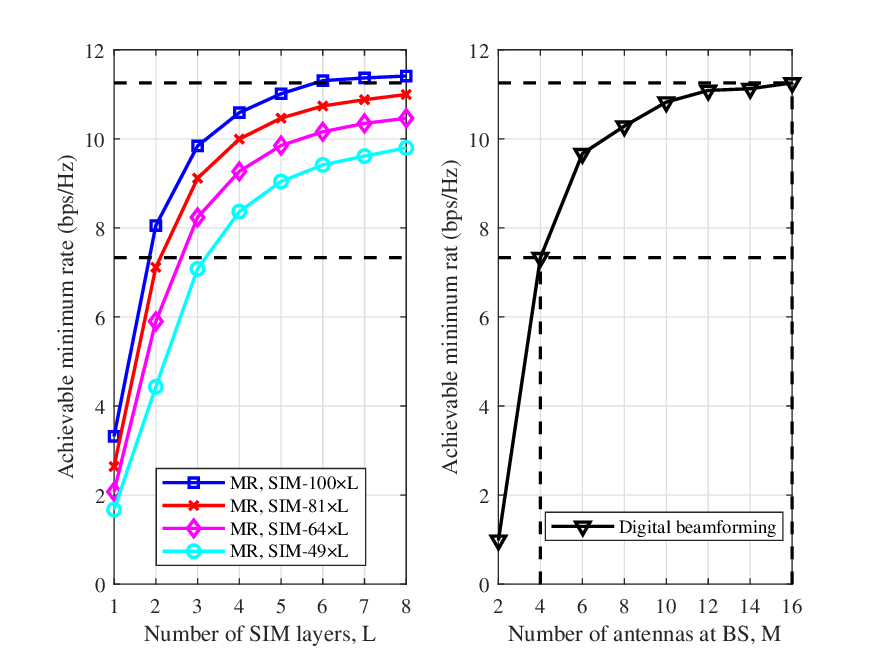}
		\label{PlotLayerMaxmin}
	}%
	\\
	\subfigure[GMR maximization.]{
		\includegraphics[width=0.44\textwidth]{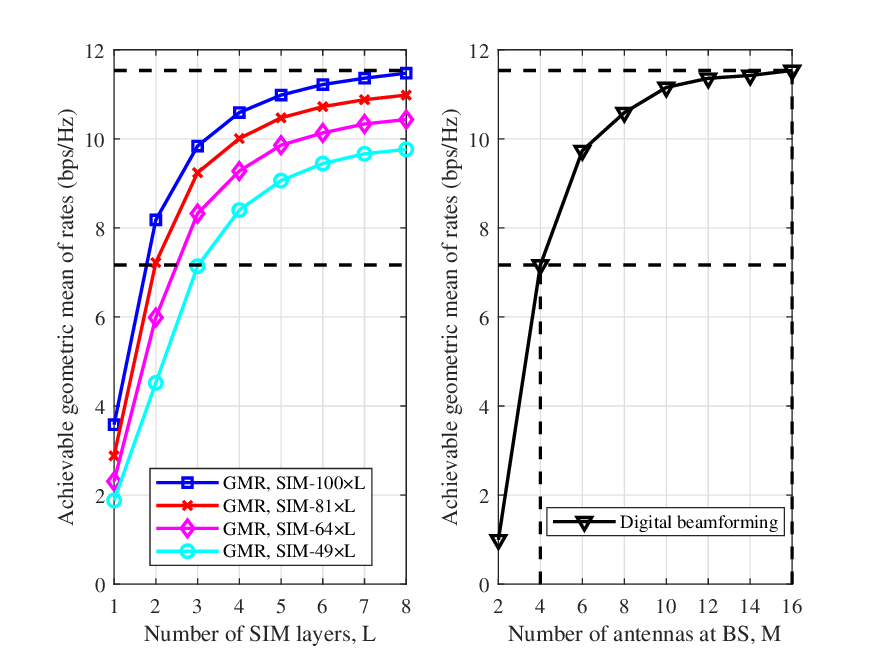}
		\label{PlotLayerGMate}
	}
	\centering
	\caption{Comparison of SIM beamforming and digital beamforming.}
	\label{PlotLayer}
\end{figure}


Fig.~\ref{PlotLayer} compares the user rate performance of the proposed SIM beamforming schemes (the left subplot) and digital beamforming (DB) scheme without SIM (the right subplot) in multiuser MISO downlink systems, using MR and GMR as performance metrics, respectively. Notably, in the considered schemes, the BS is equipped with only 4 transmit antennas (i.e., 4 RF chains). In contrast, for the DB scheme, the number of antennas $M$ is equal to the number of the required RF chains.  For DB for MR maximization and GMR maximization, we adopt the algorithms from \cite{Fang2024} and \cite{Yu2022}, respectively. Firstly, in SIM beamforming scheme, we can see that increasing  
$L$ or $N$ enhances rate performance. While in DB scheme, increasing the number of transmit antennas can also enhance the rate performance. DB reaches its performance limit at $M=16$, whereas the SIM beamforming schemes approach their rate limits at $L=8$ under different meta-atom configurations. Since the proposed schemes solely rely on power allocation without phase adjustment at BS, DB with $M=4$ outperforms SIM beamforming with $L=1$. However, SIM beamforming with $L=2$ and $N=100$ surpasses DB at $M=4$, and all SIM beamforming schemes with $L=3$ outperform DB with $M=4$. Interestingly, SIM beamforming with $L=6$ and $N=100$ achieves a similar rate as DB with $M=16$, which reveals that we can adopt SIM to replace multiple antennas in terms of rate performance, as suggested in \cite{An2023, An2023a, An2024, An2024a}. For a fair comparison, we also evaluate the power consumption of both beamforming schemes. The total power consumption is given by $P_{tol} = P_{\max} + M P_{RF} + P_0 + L N P_{SIM}$, where $P_{RF} = 30$ dBm is the circuit power consumption of each RF chain, $P_0 = 40$ dBm is the BS base power consumption, and $P_{SIM} = 10$ dBm is the power consumption per SIM meta-atom \cite{Perovic2024}. Assuming $P_0$ is the same for all schemes, SIM beamforming with 100$\times$6 consumes 20.1W, lower than the power 26.1W required by DB at $M=16$. 
\section{Conclusion}
In this paper, we investigated rate fairness optimization for SIM-assisted multiuser systems through two optimization problems: MR maximization and GMR maximization.  We gave two low-complexity algorithms based on closed-form solutions, respectively. Numerical results confirmed the benefits of SIM deployment and validated the convergence of the proposed algorithms. It also reveals: 1) MR maximization ensures near-perfect fairness and GMR maximization achieves a balance between fairness and SR; 2) For MR and SR, the proposed algorithms remarkably outperforms the benchmark algorithms; 3) SIM beamforming can achieve comparable performance to DB while consuming less power.	
	\bibliographystyle{IEEEtran}
	\bibliography{IEEEabrv,ReferenceSIM}
\end{document}